\documentclass[apj]{emulateapj}

\usepackage{apjfonts}
\usepackage{graphicx}
\usepackage{natbib}

\newcommand{\vnn}{\mbox{N}}
\newcommand{\vn}{\mbox{\bf {n}}}

\newcommand{\vx}{\mbox{\bf {x}}}
\newcommand{\vy}{\mbox{\bf {y}}}
\newcommand{\vk}{\mbox{\bf {k}}}
\newcommand{\vq}{\mbox{\bf {q}}}
\newcommand{\vrv}{\mbox{\bf {r}}}
\newcommand{\vu}{\mbox{\bf {u}}}

\newcommand{\vv}{\mbox{\bf {v}}}
\newcommand{\tonetwo}{\theta_{12}}
\newcommand{\ctonetwo}{\cos{\theta_{12}}}

\newcommand{\dthreek}{\frac{d\vk}{(2\pi)^3}}
\newcommand{\dthreeq}{\frac{d\vq}{(2\pi)^3}}
\newcommand{\dthreeu}{\frac{d\vu}{(2\pi)^3}}
\newcommand{\dthreekrad}{\frac{(4\pi)^2\;k^2dk}{(2\pi)^3}}
\newcommand{\dthreekr}{k^2dk}
\newcommand{\parone}{\frac{\partial {\bar n}(M_1,z_1)}{\partial \delta}\biggr|_{\delta =0}}
\newcommand{\partwo}{\frac{\partial {\bar n}(M_2,z_2)}{\partial \delta}\biggr|_{\delta =0}}

\newcommand{\taud}{\dot{\tau}}

\input epsf

\begin{document}


\title{Correlation properties of the
  Kinematic Sunyaev-Zel'dovich Effect and implications for dark
  energy}


\author{Carlos Hern\'andez--Monteagudo,\altaffilmark{1} Licia Verde,\altaffilmark{1} Raul Jimenez\altaffilmark{1} and David N. Spergel\altaffilmark{2}}

\affil{}

\altaffiltext{1}{Department of Physics and Astronomy, University of Pennsylvania, 209 South 33rd St, Philadelphia, PA 19104; carloshm@astro.upenn.edu, lverde,raulj@physics.upenn.edu}

\altaffiltext{2}{Department of Astrophysical Sciences, Peyton Hall, Princeton University, Princeton, NJ 08540; dns@astro.princeton.edu}

\begin{abstract}
    In the context of a cosmological study of the bulk flows in the
  Universe, we present a detailed study of the statistical properties
  of the kinematic Sunyaev-Zel'dovich (kSZ) effect. We first compute
  analytically the correlation function and the power spectrum of the
  projected peculiar velocities of galaxy clusters.  By taking into
  account the spatial clustering properties of these sources, we
  perform a line-of-sight computation of the {\em all-sky} kSZ power
  spectrum and find that at large angular scales ($l<10$), the local
  bulk flow should leave a visible signature above the Poisson-like
  fluctuations dominant at smaller scales, while the
  coupling of density and velocity fluctuations should give much
  smaller contribution. We conduct an analysis of the prospects of
  future high resolution CMB experiments (such as ACT and SPT) to
  detect the kSZ signal and to extract cosmological information and
  dark energy constraints from it. We present two complementary methods,
  one suitable for ``deep and narrow'' surveys such as ACT and one 
  suitable for ``wide and shallow'' surveys such as SPT. Both methods 
  can constraint the equation of state of dark energy $w$ to about 5-10\%
  when applied to forthcoming and future surveys, and probe $w$ in 
  complementary redshift
  ranges, which could shed some light on its time evolution. 
  This is mainly due to the high sensitivity of the peculiar
  velocity field to the onset of the late acceleration of the
  Universe. We stress that this determination of $w$ does not rely on
  the knowledge of cluster masses, although it relies on
  cluster redshifts and makes minimal assumptions on cluster physics.
\end{abstract}


\keywords{cosmic microwave background -- large scale structure of universe}

\section{Introduction}
The new generation of ground-based high-resolution cosmic microwave
background (CMB) experiments (e.g., the Atacama Cosmology
Telescope\footnote{http://www.hep.upenn.edu/act/} [ACT;\citet{K04,fowler}] and
the South Pole Telescope\footnote{http://spt.uchicago.edu/}
[SPT;\cite{RuhlSPT}]), are designed to scan with very high sensitivity
and arcminute resolution the microwave sky. Their main goal is the
study of the thermal Sunyaev-Zel'dovich (tSZ) effect \citep{tSZ80}:
the change of frequency of CMB photons due to inverse Compton
scattering by hot electrons. Such hot electron plasma are known to be
found in clusters of galaxies, and should also be present in larger
structures, such as filaments and superclusters of galaxies. This
scattering translates into a redshift independent distortion of the CMB
black body spectrum, making the tSZ effect an ideal tool to probe the
baryon distribution in the large scales of our Universe at different
cosmic epochs.  However, this is not the only effect of an electron
plasma on the CMB radiation. If a cloud of electrons is moving with
some bulk velocity with respect to the CMB frame, then Thomson
scattering by these electrons will imprint new (Doppler induced)
temperature fluctuations on the CMB photons.  This is the so called
kinematic Sunyaev-Zel'dovich effect (kSZ, \citet{kSZ72}), which is
spectrally indistinguishable from the intrinsic CMB temperature
fluctuations.

Although the kSZ effect is typically an order of magnitude smaller
than the tSZ in clusters of galaxies (and for this reason much harder
to detect), it encodes precious cosmological information since it
depends on the peculiar velocity field. Indeed, kSZ measurements can
yield valuable information about the large-scale velocity field, the
evolution of the dark matter potential, and the growth of
fluctuations. The study of large scale velocity fields (cosmic flows)
has been an active research area in the nineties.  There have been
numerous attempts to measure bulk flows using the large-scale
distribution of galaxies and their peculiar velocities, and to place
constraints on the matter power spectrum or the Universe matter
density, (see reviews of \citet{sw95} and \citet{CourteauDekel} and
references therein).  However, it became clear that these measurements
had to be corrected for systematic errors, such as the biases
introduced when calibrating the distances of the galaxies under
study, or the non linear components of the velocities of those
objects.  With kSZ observations, by using clusters as tracers of the
velocity fields, one is more confident to probe larger (less
non-linear) scales.

While there have been no kSZ detections to date, upper limits on the
peculiar velocities of individual clusters have been reported by
\cite{Bensonetal}.  Such a difficult measurement could in principle be
hampered by other effects such as non-linearities and the complicated
physics of the intra-cluster gas. \citet{Nagai+2003} have shown that
the kSZ is not diluted by the internal velocity dispersion in the
intracluster gas. \citet{MaFry2002} calculated the temperature
fluctuations produced by the kSZ in the non-linear regime using the
halo model. \citet{Bensonetal} showed that the signal-to-noise of a
kSZ measurement should be distance independent and suggested combining
signals from different redshifts.
\citet{Holder2002} and \citet{nabila} have discussed how to extract
the kSZ signal from maps, while \citet{spirou} used N-body simulations
to build templates of kSZ maps in the context of the Planck mission, and
\citet{oldjefes} study the possibility of extracting the kSZ dipole from
CMB surveys covering a large fraction of the sky.
Due to its weak signal, indirect detection of the kSZ has been
proposed through cross-correlation techniques with weak-lensing
\citep{Dore+2004} or old galaxies \citep{Dedeo+05}.  As we
shall see below, a measurement of the kSZ effect would be very
valuable, since it would not only allow us to measure bulk flows of
clusters of galaxies and test the predictions of the standard model,
but also provide additional constraints on
cosmological parameters, especially on the equation of state of dark
energy.
 
In this paper we compute the correlation function and the power
spectrum of the kSZ effect. This requires modeling of the peculiar
velocity field and the cluster population. We explore the prospects
for future CMB experiments to measure the kSZ correlation function,
and find that ACT-like experiments should be able to detect the
kSZ-induced CMB variance at high ($\sim 12\sigma$) significance level.
Further, we study the dependence of the kSZ correlation function on
the cosmological parameters, and show that it can be used to measure
the equation of state of dark energy ($w$) {\em if} the redshifts of
the clusters of galaxies detected in CMB surveys are available.  We
find that, with the SALT follow up of ACT data or with a wider but 
shallower (SPT-like) survey, the $w$ parameter can
be constrained with an accuracy of 8\% for an ACT scan of 400 square
degrees. This error should scale inversely with the square root
of the covered area and hence becomes 5\% for a 1,000 square degree area.  
These determinations of $w$ do not rely on the
knowledge of cluster masses, but do rely on clusters
redshifts and reasonable assumptions on cluster physics.


Unless otherwise stated,
throughout this work we shall asume a LCDM cosmological model \citep{WMAP}
with $\Omega_m = 0.3$, $\Omega_{\Lambda}=0.7$, $h=0.72$, and
$\sigma_8=0.88$.  The paper is organised as follows: in Section 2 we
study the statistical properties of the projected peculiar velocity
field and provide an analytical expression for the correlation
function of projected velocities. In Section 3 we compare the kSZ and
tSZ effects, and discuss the strategy to enhance the probabilities of
detecting the former. In Section 4 we compute the correlation function
and the power spectrum of the kSZ, both when we consider only a given
set of clusters present in a survey or the whole celestial sphere. In
Section 5 we outline two methods to estimate the kSZ effect in future
CMB cluster surveys and in Section 6 we explore the dependence of kSZ
measurements on cosmological parameters making particular emphasis on
$w$. We conclude in Section 7. \\

\section{The Correlation Function of Line-of-Sight Linear Peculiar Velocities}

While the measurement of the kSZ effect of individual clusters is
difficult (e.g.,\citet{aghanim01,Bensonetal}), in this paper we address the
prospects for statistical detection of cluster peculiar velocities in
future CMB surveys.  This requires the knowledge of the ensemble
properties of the velocity field traced by the galaxy cluster
population, to which we devote the current section. For clarity and
future reference, a statistical description of the linear velocity
field and related quantities is given in Appendix A. \\

Throughout this paper, we shall assume that the measured velocity
field obeys linear theory. However, as noted by \citet{colberg00},
this is not completely fulfilled by cluster velocities, since clusters
are peculiar tracers of the large scale matter distribution and show
{\em biased} velocities compared to the expectations provided by the
linear theory. \citet{colberg00} found that this bias was typically a
30\% - 40\% effect. Although several attempts have been made to
model this boost in terms of the underlying density field
\citep{sdiaferio,hamana03}, in subsequent sections it will be
accounted for by simply increasing the cluster velocities by a factor
$b_v = 1.3$ \citep{sdiaferio}.
The goal of this  paper
is to present a theoretical calculation of the detectability of the signal and the forecasted signal-to-noise: detailed comparison with
numerical simulations are left to future work.  Full treatment can be implemented only
with the help of numerical simulations  matched to a given observing program(see \citet{peel05} for a recent study).   We must stress that,
although some modelling of non-linear effects must be included in this
study, clusters are the largest virialised structures known in the
Universe, and probe much bigger scales than galaxies. Therefore we
must expect them to be significantly better tracers of the linear
velocity field. Furthermore, as we shall see below, the peculiar
velocity estimator (the kSZ effect) does {\em not} depend of distance,
which avoids the need to use redshift-independent distance indicators,
as opposed to other peculiar velocity estimators.\\


On large, linear scales, the density and peculiar
velocities are related through the continuity
equation:  $\partial \delta_{\vk} / \partial t = -i \vk\cdot\vv_k/a$,
where $a$ and $k$ are the scale factor and comoving Fourier mode,
respectively.  The peculiar velocity of a cluster, as probed by its
kSZ effect, can be interpreted as the linear peculiar velocity field
smoothed on comoving scale $R$ which corresponds to the cluster's mass
$M$ via
\begin{equation}
R = \biggl[ \frac{3 M}{4\pi {\bar \rho}}\biggr]^{1/3},
\label{eq:rscale}
\end{equation}
where ${\bar \rho}$ is the background matter density.  The
kSZ effect is sensitive to the line-of-sight component of the
velocity, but under the assumption that the velocity field is Gaussian
and isotropic (which should be satisfied in the linear regime) the
three spatial components of the velocity field must be statistically
independent. Moreover, the power spectrum must completely determine the
statistical properties of the velocity field.  Thus, in a given
cosmological model, the linear velocity field power spectrum (which in
turn is related to the matter power spectrum) should univocally
determine the angular correlation function (and angular power
spectrum) of the line-of-sight cluster velocities.\\

In linear theory, the velocity dispersion smoothed over spheres of
comoving radius $R$ (corresponding to a given cluster mass $M$) is
given by
\begin{equation}
\sigma_{vv}^2 (R,z) = \biggl( H(z)\; \biggl| \frac{d{\cal D}_{\delta}}{dz} 
                                               \biggr|\biggr)^2
        \int dk \;k^2\; \frac{P_m(k)}{2\pi^2 k^2} \bigl| W (kR) \bigr|^2,
\label{eq:vel1}
\end{equation}
where $W(kR)$ is the Fourier transform of the top hat window function,
$H(z)$ is the Hubble function, ${\cal D}_{\delta}(z)$ is the linear
growth factor and $P_m(k)$ is the present day linear matter power
spectrum, (the power spectra at any redshift will be denoted as
$P_{\delta \delta} (z,k) \equiv {\cal D}_{\delta}^2(z) \;P_m(k)$).
Hence the power spectrum of the velocities is:
\begin{equation}
P_{vv} (k) = \biggl( H(z)\; \biggl| \frac{d{\cal D}_{\delta}}{dz} 
            \biggr| \biggr)^2 \;   \frac{P_m(k)}{k^2} = 
      {\cal D}_{v}^2 \frac{P_m(k)}{k^2},
\label{eq:psv}
\end{equation}
where ${\cal D}_{v}\equiv H(z)d{\cal D}_{\delta}/dz$ is the velocity growth
factor. When computing ${\cal D}_{\delta}$ for different Dark Energy
models, we used the analytical fit provided by \citet{linder05}:
\begin{equation}
g(a) = \exp{ \int_0^a\;d \log a \;
  \biggl[\bigl(\Omega_m\frac{H_0^2}{a^3H^2(a)}\bigr)^{\gamma} -1\biggr]},
\label{eq:gfwrel}
\end{equation}
where $g(a)\equiv {\cal D}_{\delta}(a) / a$ gives the deviation of the
growth factor from that of a critical ($\Omega_m = 1$) universe, and $\gamma$
is given by 
\begin{equation}
\gamma = 0.55 + b[1+w(z=1)],
\label{eq:gammaexp}
\end{equation}
with $b=0.05$ if $w>-1$ and $b=0.02$ otherwise.\\

\begin{figure}
\begin{center}
         \epsfxsize=8cm \epsfbox{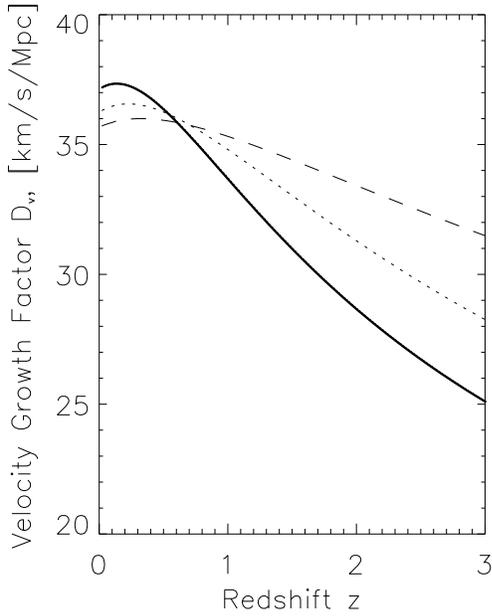}
\caption[fig:Dv]{Peculiar velocity growth factor for three different
cosmologies: a $\Lambda$CDM model ($\Omega_L=0.7$, $\Omega_m=0.3$,
thick solid line), a $\Omega_m=0.3$, flat
$w=-0.6$ model (dotted line), and  $\Omega_m
= -1/3$ and $w=-1/3$, (dashed line).  }
\label{fig:Dv}
\end{center}
\end{figure}

The growth of the velocity perturbations with redshift may provide
useful cosmological constraints such as constraints on the equation of
state of dark energy \citep{Dedeo+05}. This is illustrated in
Fig.(\ref{fig:Dv}), where we show the redshift evolution of the
velocity growth factor for three different cosmological models: a
$\Lambda$CDM model ($\Omega_{\Lambda}=0.7$, $\Omega_m=0.3$, thick
solid line), a flat universe with $\Omega_m = 0.3$ and dark energy
equation of state parameter $w=-0.6$ (dotted line), and another flat
model with $\Omega_m = 0.3$ and $w=-1/3$, (dashed line).
Due to the $k^2$ factor in the denominator of eq.(\ref{eq:psv}), the
signal is weighted by the largest scales, making this probe relatively
insensitive to the smoothing scale and therefore to clusters mass.
Indeed, if the dependence of $\sigma_{vv}$ versus mass is approximated
by a power law, then one finds that for the concordance
model $\sigma_{vv} \propto M^{-0.13}$.\\

Having this in mind, we compute here the angular correlation function
of the {\em line-of-sight} (LOS) cluster velocities.  Note that  
\citet{peel05} takes a different approach to this calculation. Assuming
that we can measure the LOS component of the peculiar velocity of a
cluster, we compute the quantity
\begin{equation}
C_{vv} (\theta_{12}) \equiv \langle \biggl(\vv (\vx_1) \cdot \vn_1 \biggr) 
            \biggl( \vv (\vx_2) \cdot \vn_2\biggr) \rangle,
\label{eq:corv1_tex}
\end{equation}
where $\vn_1$ and $\vn_2$ denote two different directions in the sky,
``connecting'' the observer to the cluster positions $\vx_1$ and
$\vx_2$ and $\theta_{12}$ denotes the angle between $\vn_1$ and
$\vn_2$.  We refer the reader to Appendix B for the detailed
derivation and here we report the final expression for this
correlation function:
\[
 C_{vv} (\theta_{12}) = 
\sum_{even\; l} \frac{2l+1}{4\pi} \; \cos{\theta_{12}} \times 
           \phantom{xxxxxxxxxxxxxxx}
\]
\[
\phantom{xxxx} 
          \biggl( \frac{2}{\pi}\; {\cal F}_l \biggr) \int k^2dk\; 
P_{vv} (k) \;
           W(kR_1)\; W(kR_2) \; \times \; 
\]
\begin{equation}
\phantom{xxxx} 
j_l (k[x_1 - x_2\cos{\theta_{12}}]) 
           \; j_l (k x_2 \sin{\theta_{12}}),
\label{eq:corv2_tex}
\end{equation}
In this equation, the factor ${\cal F}_l$ is given by 
\begin{equation}
{\cal F}_l \equiv \frac{(l-1)!!}{2^{l/2} \; 
  (l/2)!} \;\; \cos{l\pi/2},
\label{eq:F_l_tex}
\end{equation}
and $x_1, x_2$ are the (comoving) distances to the clusters, (without
loss of generality we have used the convention that $x_1 \geq x_2$).
Here $j_l(x)$ denote the spherical Bessel functions and the summation
must take place only over {\em even} values of $l$; $R_1$ and $R_2$
refer to the linear scales corresponding to the masses of each
cluster.  Note that we recover eq.(\ref{eq:vel1}) in the limit of
$\theta_{12}
\rightarrow 0$ and $x_1 \rightarrow x_2$.\\

\begin{figure}
\begin{center}
         \epsfxsize=8cm \epsfbox{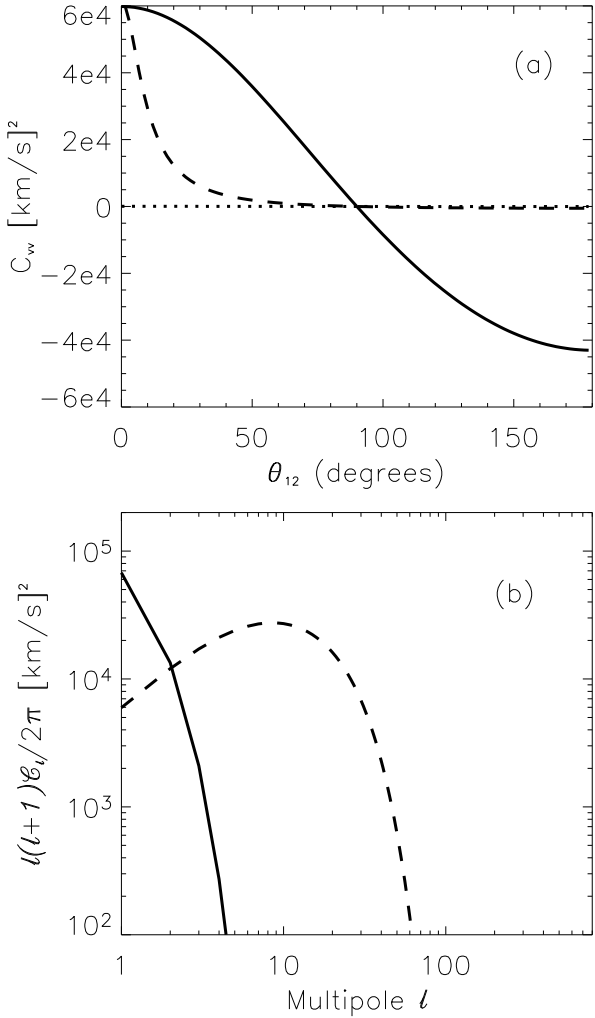}
\caption[fig:cvv1]{Correlation function (top) and 
  corresponding power spectrum (bottom) of the projected peculiar
  proper velocities. The solid lines observe the case where the two
  clusters are placed at $z=0.005$ from the observer.  Dashed lines
  show the case in which clusters are further, $z=0.01$. Finally, the
  dotted line considers the case where one cluster is located at
  $z=0.1$ and the other cluster is much further ($z=1$). Note the lack
  of correlation in this case.}
\label{fig:cvv1}
\end{center}
\end{figure}

Fig.(\ref{fig:cvv1}a) shows the behaviour of $C_{vv}$ vs $\tonetwo$ in
the concordance $\Lambda$CDM model for a couple of $10^{14}$
M$_{\odot}$ clusters when they are both placed at z=0.005 (solid
line), both placed at z=0.1 both (dashed line) and when once cluster
is at z=0.1 and the other at z=1 (dotted line).  In the first (clearly
unrealistic) case, the clusters are so close to us that both are
comoving in the same bulk flow with respect to the CMB frame, giving
rise to the dipolar pattern shown by the solid line. In the case where
both clusters are at $z\sim 0.1$ (dashed line), their correlation
properties are strongly dependent on $\tonetwo$, since this angle
defines the distance between the clusters. For small angular
separation, the clusters are still relatively nearby, and hence their
peculiar velocities are correlated, but this correlation dies as the
separation of the clusters increases. The angular distance at which
the correlation drops to half its value at zero separation is $\sim$
10\degr, corresponding to roughly 40 $h^{-1}$ Mpc. Finally, if
clusters are very far apart from each other (thick dotted line), their
peculiar velocities are not correlated.\\

An alternative way to present the correlation properties of the
projected peculiar velocities of the clusters is the velocity field
angular power spectrum.  In  Appendix B, we invert $C_{vv}
(\tonetwo)$ into its angular power spectrum $C^{vv}_l$, i.e.,
\begin{equation}
C_{vv} (\tonetwo ) = \sum_l \frac{2l+1}{4\pi} C^{vv}_l P_l(\cos{\tonetwo} ).
\label{eq:cvv_cls}
\end{equation}
We find that 
\begin{equation}
C^{vv}_l \equiv  \frac{4\pi}{2l+1} \bigl(l{\cal B}_{l-1} + (l+1){\cal B}_{l+1} 
            \bigr), 
\label{eq:cls_tex}
\end{equation}
where the ${\cal B}_{l}$ coefficients are defined by
\begin{equation}
{\cal B}_l \equiv  4\pi \int \frac{k^2dk}{(2\pi)^3} P_{vv}(k)\; j_l(k x_1)
        j_l(k x_2)
\label{eq:als_tex}
\end{equation}
Fig.(\ref{fig:cvv1})b displays the power spectra for two cases: two
very nearby clusters (both at z=0.005, solid line), showing an almost
dipolar pattern, and two relatively far away clusters (both at z=0.1,
dashed line), case in which the power is transferred to higher
multipoles. The power spectrum for the clusters placed at z=0.005 and
z=0.1 is zero.\\

Here we have characterised the projected peculiar velocity field at
cluster scales.  Next, we address the study of the kSZ effect and its
comparison with the tSZ effect.

\section{The Sunyaev-Zel'dovich Effects}

\subsection{The kinematic Sunyaev-Zel'dovich Effect}

The Kinematic Sunyaev-Zel'dovich effect describes the Doppler kick
that CMB photons experience when they encounter a moving cloud of
electrons. Since this is simple Thomson scattering, there is no
change of the photon frequency, and hence it leaves no spectral
signatures in the CMB blackbody spectrum. Therefore, this effect is
solely determined by the number density of free electrons and their
relative velocity to the CMB frame. An observer will only be sensitive
to the {\em radial} component of the electron peculiar velocity, so
the expression for the change in brightness temperature becomes
(Sunyaev \& Zel'dovich 1968)
\begin{equation}
\frac{\delta T_{kSZ}}{T_0} = \int dl\; n_e(l)\; \sigma_T\; 
  \biggl( - \frac{{\bf v}\cdot {\bf n} }{c}\biggr) \equiv
 \tau \; \biggl( - \frac{{\bf v}\cdot {\bf n}}{c}\biggr).
\label{eq:dtkSZ1}
\end{equation}
 
Here, we have assumed that the peculiar velocity is the same for
all electrons. $\tau$ stands for the optical depth, and ${\bf n}$ is a unitary
vector giving the direction of observation.  This process must take
place in two contexts:  (1) When the intergalactic medium
becomes ionized by the high energy photons emitted by the first stars,
inhomogeneities in the electron velocity and density distributions generate
a kSZ signal which is known as the Ostriker-Vishniac effect.
Despite of the large size of ionized structures encountered by the CMB
photons, the electron density contrast is relatively small, and
further, we do not know where in CMB maps to look for this signal
because we do not know the location of the ionized bubbles which formed
during reionization. The amplitudes and angular scales at which this
effect should be visible are model dependent, but can be as high as
a few microK in the multipole range $l>2000$, \citep{santos}.
(2) Clusters of galaxies at lower redshift leave a more
easily detectable signal. Their high electron density can give rise to
values of $\tau$ as high as $10^{-3}-10^{-2}$, and their peculiar
velocities should be close to 300 kms$^{-1}$ at $z=0$, which together
can produce temperature fluctuations of the order of 1-10 $\mu$K.  The
clusters high optical depth is also responsible for large spectral
distortions of the CMB generated through the thermal
Sunyaev-Zel'dovich (tSZ) effect. The tSZ effect is typically an order
of magnitude larger than the kSZ and introduces {\em frequency
  dependent} brightness temperature fluctuations which, in the not
relativistic limit, change sign at 218 GHz. Therefore by combining
observations in bands at frequencies lower and higher than this cross
frequency, it is possible to obtain the cluster position and to
characterize the tSZ cluster signal.  Once the cluster position is
identified their kSZ contribution should be accessible at $~$218
GHz. However, as noted by \citet{neelima} and earlier by \citet{holder04},
even with measurements in three different frequencies it may not be possible
to obtain a clean estimate of the kSZ effect for a single cluster. We shall
show below that we are not interested in a very accurate kSZ estimate for
a given cluster, but on {\em unbiased} estimates on our entire cluster sample.
\\

In the next subsection we make a detailed comparison of the amplitude
of the kSZ and the tSZ effects.

\subsection{Comparison of the kSZ and the tSZ effects in clusters of galaxies}

In what follows we shall describe the galaxy cluster population by
adopting the model presented in \citet{vhs}. This model is based upon
the spherical collapse description of galaxy clusters
\citep{sphcol}, and assumes that clusters are isothermal and their
gas acquires the virial temperature of the halo. The halo mass and
redshift distribution is approximated by the formalism presented in
\citet{ST99}. We refer to \citet{vhs} for further details in this
modelling.\\

Both kSZ and tSZ effects can be written as integrals of some function
${\cal K}(r)$ along the line of sight crossing the cluster, weighted
by the electron density:
\begin{equation}
\frac{\delta T}{T_0}=g(x)\int dl\; \sigma_T\; n_e(l) {\cal K}(l)
\end{equation}
For the non-relativistic 
tSZ $g(x)=\bigl( x\; \coth{x/2} - 4 \bigr)$, $x\equiv h\nu
/k_BT_0$ is the adimensional frequency in terms of the CMB monopole
$T_0$, ${\cal K}=k_BT_e/(m_e c^2)$ where $n_e$, $T_e$, $m_e$ are the
electron density, temperature and mass respectively, $\sigma_T$ the
Thomson cross-section and $k_B$ the Boltzmann constant.  For the kSZ,
$g(x)\equiv 1$, ${\cal K}=-\vn\cdot\vv/c$, where $\vn$ is a unitary
vector pointing along the line of sight, and $\vv$ is the cluster
peculiar velocity.  If clusters are perfectly virialised objects, then
one must expect the scaling $T_e \propto M^{2/3}$. However, following
\citet{vhs}, the cluster model should leave some room for some
deviations from such scaling, which can be due to internal
(non-linear) cluster physics, deviations from purely gravitational
equilibrium, preheating, etc. Our parametrization adopts $T_e \propto
M^{1/\xi}$, where $\xi=1.5$ for a perfectly virialized cluster, but
spans the range from 1.5 to 2 in the literature. We shall present
results for $\xi=1.5$, but the sensitivity of our results on $\xi$ is
minimal. At the same time, the radial peculiar velocity dispersion of
the clusters, $\langle \sqrt{ ( -\vn\cdot\vv/c )^2 }\rangle =
\sigma_{vv} $ decreases very slowly with mass ($\sim M^{-0.13}$), so,
at fixed redshift we must expect the ratio kSZ/tSZ to be bigger for
low mass clusters, which are more numerous. Regarding the redshift
dependence, clusters tend to be denser at earlier epochs (the product
$n_e\;r_v$ scales roughly as $(1+z)^2$), so we must expect larger tSZ
and kSZ amplitudes at higher redshifts. However, the dependence of the
function ${\cal K}$ is different in each case: while clusters are
hotter at earlier times ($T_e \propto (1+z)$), velocities have not had
so much time to grow as at present epochs ($\sigma_{vv} \propto
1/\sqrt{1+z}$), and hence the ratio kSZ/tSZ ($\propto 1/(1+z)^{3/2}$)
decreases with redshift.\\

\begin{figure}
\begin{center}
        \epsfxsize=8cm \epsfbox{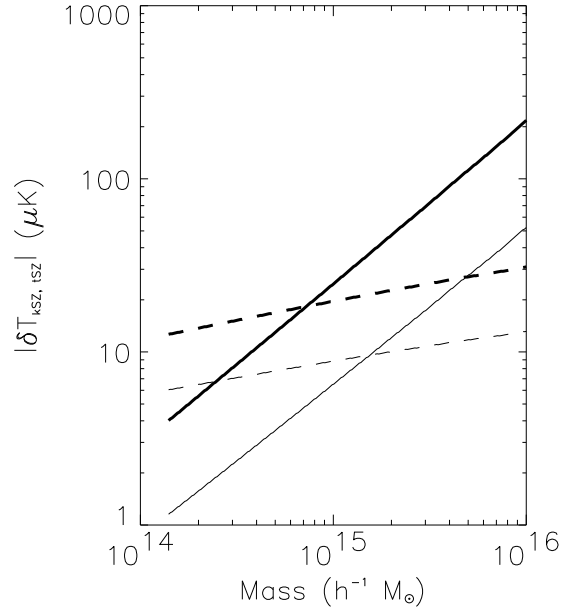}
\caption[fig:tSZ_kSZ]{Comparison of the tSZ (solid lines) and kSZ 
(dashed line) effects  as a function of clusters mass. Thick lines refer
to z=1 (where most of the clusters are located for ACT-like surveys), 
whereas thin lines correspond to z=0. For the majority of clusters, the
kSZ flux will be a few times bigger than the tSZ flux at 222 GHz.
}
\label{fig:tSZ_kSZ}
\end{center}
\end{figure}

We show these scalings explicitely in Fig.(\ref{fig:tSZ_kSZ}), where
the thin lines evaluate the tSZ and the kSZ at $z=0$ and thick lines
correspond to $z=1$.  The solid lines provide the amplitude of the
cluster-induced tSZ fluctuations at 222 GHz, which, as we shall see
below, will be taken as our effective frequency after accounting for
the tSZ relativistic corrections and the effects related to the finite
spectral width of the detectors.  The dashed lines provide the
expected rms amplitude of the kSZ effect in clusters: we see that, at
$z=0$, clusters of masses below $\sim 10^{15}$ $M_{\odot}$ should
produce more kSZ than tSZ flux at 222 GHz, at least by a factor of a
few. This mass threshold decreases at $z=1$, but so does the typical
halo mass at such redshift (clusters of $\sim 10^{14}\; M_{\odot}$ are
very rare objects), so again we must expect the kSZ to dominate over
the tSZ.

\section{The power spectrum of the kSZ effect}

\subsection{Model of the cluster population}

We next study the kSZ signal generated by the entire population of
clusters of galaxies by computing its two second-order momenta, i.e.,
the correlation function and the power spectrum.  For this, it is
first necessary to have a model to describe the population of galaxy
clusters in the Universe. We shall adopt the hierarchical scenario in
which small scale overdensities in the Universe become non-linear and
collapse first, and then merge and give rise to bigger non-linear
structures. The abundance of haloes of a given mass at a given cosmic
epoch or redshift is given by the clusters mass function. We will
adopt the Sheth \& Tormen (hereafter ST) mass function \citep{ST99},
which will be denoted by $\bar{n}(M,z) \equiv dN(M,z)/dV(z)$ and provides
the {\em average} number density of haloes of mass contained between
$M$ and $M+dM$ at redshift $z$. It must not be confused with
  $n(M,\vx)$, which is the {\em actual} number of haloes in that mass
  range at position $\vx$. The latter can be understood as a {\em random}
  variable, the former as its {\em mean}. As it will be useful later, we
first compute the mean number of haloes present in two volume elements
centered at $\vx_1$ and $\vx_2$:
\[
\langle n(M_1, \vx_1) n(M_2, \vx_2) \rangle = 
   \bar{n}(M_1, z_1)  \bar{n}(M_2,z_2)  \; +
\]
\[
\phantom{xxxx}
 \delta_D (M_1-M_2) \delta_D^3 (\vx_1 - \vx_2) \bar{n}(M_1, z_1) 
  \;+\;
\]
\begin{equation}
\phantom{xxxx}
 \langle \Delta [ n(M_1, \vx_1) ] \Delta [n(M_2, \vx_2) ] \rangle.
\label{eq:corr_sources}
\end{equation}
In this equation, $z_1$ and $z_2$ are the redshift corresponding to
positions $\vx_1$ and $\vx_2$ respectively. The first term in the
right hand side of the equation is merely a constant, but will have
its relevance, because it will couple with the velocity field as we
shall see below.  The next term containing the Dirac deltas accounts for
the (assumed) Poissonian statistics ruling the (discrete) number density of
sources, and will be referred to as the {\em Poissonian} term. In the
third term, $\Delta [n(M_1,\vx_1)]$ stands for the deviation with
respect to the average halo number density due to the environment,
i.e., due to large scale overdensities, which condition the halo
clustering. Therefore, this third term describes the spatial
clustering of haloes, which is a biased tracer of the spatial
clustering of matter.  In the extended Press-Schecter approach it can
be shown that the power spectrum and the correlation function of
haloes and underlying matter are merely proportional to each other
over a wide range of scales.  This is commonly expressed by a {\em
  bias factor} \citep{MoW96}, so that
\begin{equation}
\xi_{hh} (r) = b^2 (M,z) \;\xi_m (r)\; ; \;\; P_{hh}(k) = b^2 (M,z) P_m(k).
\label{eq:bias}
\end{equation}

Here $\xi_{hh}, \xi_m$ and $P_{hh}(k), P_m(k)$ stand for halo-halo and
matter correlation function/power spectrum, respectively. Therefore, the
third term in the RHS of eq.(\ref{eq:corr_sources}) equals 
$\bar{n}(M_1,z_1)\bar{n}(M_2,z_2)b(M_1,z_1)b(M_2,z_2)\xi_m (\vx_1-\vx_2)$.\\

A parallel approach consists in writing the halo mass function as a
function of some linear scale matter overdensity $\delta \equiv (\rho
- {\bar \rho}) / {\bar \rho}$.  The number of haloes at $\vx$ is then
approximated as
\begin{equation}
n(M,\vx) = \bar{n}(M,z) + \eta + \frac{\partial \bar{n}(M,z)}{\partial \delta}
     \biggr|_{\delta = 0} \delta (\vx )  + {\cal O} [\delta^2],
\label{eq:nclcond}
\end{equation}
where $\eta$ is a random variable which introduces the Poissonian
behaviour of the source counts. In the extended Press-Schechter
formalism, it turns out that $\partial \bar{n}(M,z)/\partial
\delta\bigr|_{\delta = 0} $ coincides with the bias factor $b(M,z)$,
and by using this it is possible to reproduce eq.
(\ref{eq:corr_sources}) from eq.(\ref{eq:nclcond}).  This justifies
neglecting all higher-order powers of $\delta$ in eq.
(\ref{eq:nclcond}). We have taken $\delta$ to be in {\em linear}
regime, but, since $\xi_{hh}(r) \simeq b^2 \xi_{mm}(r)$ down to scales
comparable to the halo size, we shall use this formalism down to halo
scales.

\subsection{A Line of Sight approach for the kSZ effect}

We next write the temperature anisotropies induced by the kSZ in
clusters of galaxies as an integral along the line of sight:
\begin{equation}
\frac{\Delta T_{kSZ}}{T_0}[\vn ] =\int_0^{r_{lss}} dr \;
  \sum_{j} \dot{\tau}_j \; \biggl( - \frac{{\bf v_j}\cdot {\bf n}}{c}\biggr)
               W^{gas}_j (\vrv - \vrv_j).
\label{eq:dtksz1}
\end{equation}
Here, $r$ is the comoving radius integrated to the last scattering
surface, and the sum over the index $j$ represents a sum on all
clusters; $\dot{\tau}_j$ denotes the opacity in the {\em center} of
the cluster ($\taud=a\sigma_T n_{e,c}$ with $n_{e,c}$ the central
electron number density and $a$ the scale factor), and the window
function $W^{gas}_j(\vrv - \vrv_j)$ denotes the gas profile of the
cluster. Although we should adopt some realistic shape for this
profile \citep{KS01}, we have adopted a simple Gaussian window with
scale radius equal to the virial radius of the cluster. This is
justified since, in Fourier domain, at scales bigger than the cluster
size, the window function is merely equal to the volume occupied by
the gas, regardless of the shape of the gas profile.  This step
simplifies our computations significantly, {\em and} does not
compromise the accuracy in the relatively big scales (bulk flow
scales) in which we are interested.  Note that even if the sum is made
over all clusters, {\em only} the clusters being intersected by the
line of sight will contribute to the integral. This sum can be
re-written first as an integral and then as a convolution,
\[
\frac{\Delta T_{kSZ}}{T_0}[\vn ] =\int_0^{r_{lss}} dr \;\int dM\; \times
 \]
\[
\phantom{x}
\biggl( \int d\vy \;
\dot{\tau}(M,z) \; \biggl( - \frac{{\bf v(M,\vy)}\cdot {\bf n}}{c}\biggr)\;n[M,\vy] 
                        \; W^{gas}\bigl[ M, \vrv - \vy \bigr] \biggr)
\]
\[
\phantom{x} = \int_0^{r_{lss}} dr \;\int dM\; \times
 \]
\begin{equation}
\phantom{x}
\biggl[ 
\bigl( \dot{\tau}(M,z) \; \biggl( - \frac{{\bf v}(M)\cdot {\bf n}}{c}\biggr)
\;n\bigl[ M \bigr] \bigr)
                   \star W^{gas}\big[M,z\bigr] \biggr] [\vrv ].
\label{eq:dtksz2}
\end{equation}
Note that central optical depth and the window function have an
intrinsic dependence on redshift. This is due to the fact that
clusters formed at higher redshift tend to be more concentrated. The
symbol $\star$ denotes here
convolution in real space.\\

\subsection{The {\em all sky} correlation function and power spectrum of the 
kSZ effect}

If two different lines of sight are now combined to estimate the
angular correlation function, then one obtains that
\[
\biggl< \frac{\Delta T_{kSZ}}{T_0}[\vn_1 ] \frac{\Delta T_{kSZ}}{T_0}[\vn_2 ]
 \biggr> \propto
\]
\begin{equation}
\biggl< n(M_1,\vrv_1) \biggl(\vv(M_1,\vrv_1)\cdot\vn_1\biggr) 
 n(M_2,\vrv_2) \biggl(\vv(M_2,\vrv_2)\cdot\vn_2\biggr) \biggr>.
\label{eq:4term}
\end{equation}

Next we plug eq.(\ref{eq:nclcond}) and make use of the Cumulant
Expansion Theorem.  We must note as well that Poissonian fluctuations
will be assumed to be independent of $\delta$, and that for Gaussian
statistics, the 3- and 4-point functions are zero, just as $\langle
\vv \rangle$ and $\langle \delta \rangle$. Therefore we are left with
a sum of products of 2-point functions of the form

\[
 \bar{n}(M_1,z_1) \sigma_{vv}^2(M_1,z_1) \; + 
\]
\[
 \phantom{xx} \bar{n}(M_1,z_1) \bar{n}(M_2,z_2) 
   \biggl< \biggl( \vv(M_1,\vrv_1)\cdot \vn_1 \biggr)
             \biggl(\vv(M_2,\vrv_2)\cdot \vn_2 \biggr)\biggr> \;+ 
\]
\[
\phantom{xx}
 \parone\partwo \times 
\]
\[
\phantom{xx}
 \biggl[\langle \delta(\vrv_1) \delta(\vrv_2) \rangle 
    \biggl< \biggl(\vv(M_1,\vrv_1)\cdot \vn_1\biggr) 
       \biggl(\vv(M_2,\vrv_2)\cdot \vn_2 \biggr)\biggr> \; +
\]
\[
 \phantom{xx}
 \biggl< \delta(\vrv_1) \biggl(\vv(M_2,\vrv_2)\cdot \vn_2\biggr)  \biggr> 
    \biggl< \delta(\vrv_2) \biggl(\vv(M_1,\vrv_1)\cdot \vn_1  \biggr)\biggr> \; +
\]
\begin{equation} 
\phantom{xx}
   \biggl< \delta(\vrv_1) \biggl(\vv(M_1,\vrv_1)\cdot \vn_1 \biggr) \biggr> 
    \biggl< \delta(\vrv_2) \biggl(\vv(M_2,\vrv_2)\cdot \vn_2  \biggr)\biggr> \biggr].
\label{eq:array}
\end{equation}

Note that the last term is a constant.  We refer again to Appendix A
where the cross terms $\langle \delta (\vv\cdot\vn) \rangle$ are
studied. In Appendix C we provide a explicit computation of the power
spectra arising from each of the terms considered in eq.(\ref{eq:array}).
Since the last term introduces no anisotropy, the kSZ power spectrum
is the sum of four contributions: a Poisson term ($Poisson$), a term
proportional to the velocity-velocity correlation ($vv$ term), a term
proportional to the product of the velocity-velocity correlation and
the density-density correlation ($vv-dd$ term), and a term
proportional to the density-velocity squared ($dv-vd$ term).
Therefore,

\begin{equation}
C_l = C^P_l + C^{vv}_l + C^{dd-vv}_l + C^{dv-vd}_l.
\label{eq:sumcls}
\end{equation}

\begin{figure}
\begin{center}
        \epsfxsize=8cm \epsfbox{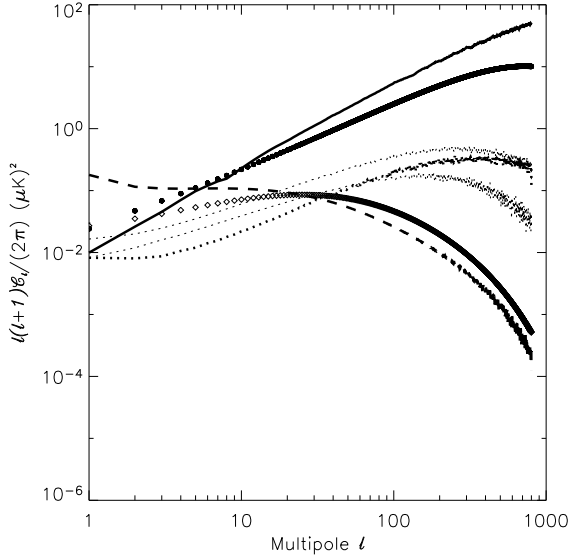}
\caption[fig:cls_as]{Different components of the all-sky kSZ power spectra: 
  the thick solid line shows the Poisson term, which dominates over
  the other terms except in the low $l$ limit. In this large scale
  range, the {\it vv} generated by the local bulk flow is the one
  introducing most power, (dashed line). Filled circles and diamonds show
  semi-analytical approximations for these two {\it Poisson} and {\it
    vv}, respectively (see eqs.(\ref{eq:p_approx},\ref{eq:vv_approx})).
  Of lower amplitude, the thick dotted line shows the sum of the {\it
    dd-vv} and the {\it dv-vd} terms, (thin dotted lines, 
note that the latter is  negative).  }
\label{fig:cls_as}
\end{center}
\end{figure}

Fig.(\ref{fig:cls_as}) displays each of the terms in
eq.(\ref{eq:sumcls}): the thick solid line is the Poisson term,
whereas the thick dashed line corresponds to the {\it vv} term.
Although the latter decreases rapidly with increasing $l$, at the
large scales it dominates over all other terms, reflecting the
presence of the local bulk flow. In an attempt to simplify the
expressions for these two terms given in Appendix C, we have found the
approximate integral
\begin{equation}
C_l^P \approx  \int dz\frac{dV(z)}{dz}\;dM\; \bar{n}(M,z)\; \sigma_{vv}^2(M,z)
\frac{\bigl| y_l (M,z)\bigr|^2}{\pi}
\label{eq:p_approx}
\end{equation}
for the Poisson term, and
\[
C_l^{vv} \approx  \int dz\frac{dV(z)}{dz} 
            P_{vv}\biggl(z,k=\frac{l}{r(z)}\biggr)\; \times
\]
\begin{equation}
\phantom{xxxxxxxxxxxx}
 \biggl[ \int\;dM\; \frac{1}{4\pi}\bar{n}(M,z)\; 
\bigl| y_l (M,z)\bigr| \biggr]^2
\label{eq:vv_approx}
\end{equation}
for the {\it vv} term. $y_l (M,z)$ is the Fourier transform of the
cluster profile in the sphere,
\begin{equation}
y_l = \biggl( \sqrt{2\pi} \theta_v \biggr)^2 \; \exp{-l(l+1) \theta_v^2 / 2},
\label{eq:yl}
\end{equation}
and $P_{vv}(k,z)$ is the $k$ and $z$ dependent velocity power
spectrum. $\theta_v$ is the angular size of the cluster virial radius.
Filled circles for the Poisson term and diamonds for the {\it vv} term
provide a comparison of these approximations with the exact integrals.
Although the amplitudes and slopes are not too disimilar, the
approximated $vv$ term is particularly innacurate at large angular
scales ($l < 10$). The approximation for the Poissonian term seems to
predict a somewhat correct amplitude at low $l$, but a shallower
slope, which translates into a smaller amplitude at high multipoles.\\

The {\it dd-vv} and {\it dv-vd} terms are shown by the thin dotted
lines. We must remark that the {\it dv-vd} term is negative, and we
are plotting its absolute value. The sum of both is given by the thick
dotted line.  The sum of these two terms is particularly hard to
detect, since it never dominates, not at large scales (it is about a
factor of 20 below the {\it vv} term), nor at small scales, where it
is well below the Poisson term.

\begin{figure}
\begin{center}
        \epsfxsize=8cm \epsfbox{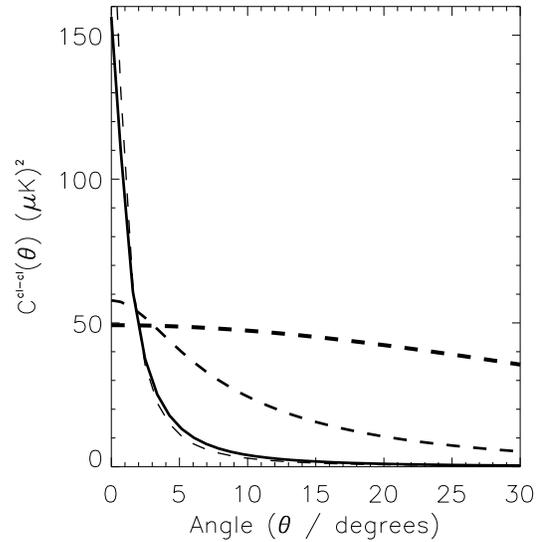}
\caption[fig:cl_cl_cf]{Cluster-cluster kSZ correlation functions for 
  a sample of clusters with masses above $2\times 10^{14}$M$_{\odot}$.
  When forming cluster pairs we require pair constituents to be at
  similar redshifts. If we consider all redshifts simultaneously, we
  obtain the solid line. The dashed lines correspond to {\it
    differential}-in-redshift cluster-cluster correlation functions:
  the thick dashed line considers only clusters around $z\sim 0.01$,
  whereas the intermediate dashed line corresponds to $z\sim 0.1$ and
  the thin dashed line to $z\sim 1$.  }
\label{fig:cl_cl_cf}
\end{center}
\end{figure}

\subsection{The {\em cluster-cluster} kSZ correlation function}

Future high resolution multi frequency CMB experiments like ACT or SPT
can provide kSZ estimates on those regions of the sky where clusters
of galaxies have been identified via tSZ.  Therefore one could attempt
to measure the velocities correlation function by computing the kSZ
correlation function in this {\em restricted} set of pixels:
\[
C_{kSZ}^{cl-cl}(\theta)=\sum_{i_1,i_2}\sum_{j_1,j_2} 
\tau (M_{i_1},z_{j_1}) \tau (M_{i_2},z_{j_2}) \; \times
\]  
\begin{equation}
\phantom{xxxxxxx}
\langle \frac{1}{c^2} \;\; \biggl(\vv (M_{i_1},z_{j_1}) \cdot \vn_1 \biggr) 
            \biggl( \vv (M_{i_2},z_{j_2}) \cdot \vn_2 \biggr) \rangle.
\label{eq:ckszcf1}
\end{equation}

Since the signal comes from clusters which velocities are correlated,
we should consider pairs close in redshift.  The solid line in
Fig.(\ref{fig:cl_cl_cf}) shows the correlation function to be measured
from {\em all} galaxy clusters from $z=0$ upto $z=4$ more massive than
$2\times 10^{14}h^{-1}M_{\odot}$, according to the standard
$\Lambda$CDM cosmology and the Sheth-Tormen mass function. Members of
the cluster pairs must be within $\Delta z= 0.01$.  Since the kSZ
amplitude per cluster is typically of few tens of microKelvin, the
zero-lag correlation function can be as high as $\sim 150$ ($\mu$K )$^
2$, and drops to one half of this value at $\theta \sim
2\degr-3\degr$. The thick dashed line shows the correlation function
for clusters located at $z\simeq 0.01$. As the redshift increases
($z\simeq 0.1$ medium thickness dashed line; $z\simeq 1$, thin dashed
line), the amplitude at zero lag increases (clusters are more
concentrated) and the correlation angle decreases.  The solid line
shows the redshift-integrated cluster-cluster correlation function:
the signal is dominated by high redshift clusters, more concentrated
and more numerous per unit solid angle.

\citet{diaferios} pointed out a potentially important non-linear
aspect that is not included in our modeling: in very massive
superclusters, the kSZ effect shows typically a dipolar pattern,
plausably caused by the encounter of two oppossite bulk flows at their
common attractor's position. Since our model predicts no dipolar
pattern at scales of a few degrees, such scenario is not accounted for
by our approach. Hence, in a realistic application the core of such
overdense regions should be excised from the analyses.  Such massive
structures however are very rare and form at very late epochs, and
their exclusion should not compromise the analysis presented here.

\section{Can the kSZ effect be measured?}

In this section we outline two different procedures to extract the kSZ
signal from future high-resolution and high-sensitivity CMB
experiments.  The procedures presented here may be sub-optimal but our
aim is to quantify the relative importance of difference sources of
error and to roughly forecast the expected signal to noise for those
experiments.  We defer the developement of an optimal procedure to
future work.  We shall try to extract the kSZ signal in a statistical
sense: while previous works (e.g., \citet{nabila}) have addressed the
difficulty of separating the kSZ effect from potential contaminants
(tSZ, radio-source emission, infrared galaxies, CMB, etc) in a given
cluster, our approach will consist on combining the signal coming from
subsets of clusters in such a way that the contribution of the noise
sources averages out.  As we shall see, the success of this procedure
will rely on the precision to which the {\em average} properties of
the potential contaminants are known.\\

The first approach, which we shall refer to as method (a), is based on
measurements of the kSZ flux and its redshift evolution. Its
sensitivity to cosmology increases with redshift, and, for this reason
it is suited for ``deep and narrow'' survey strategies.  Here we will
use ACT's specifications.  Our second approach (method (b)) uses the
ratio of kSZ and tSZ induced temperature anisotropies. This ratio is
particularly sensitive to $w$ at $z ^{<}_{\sim} 0.8$ and since cosmic
variance for the peculiar velocity field is more important at low
redshifts, this method is more suitable for ``wide and shallow''
survey strategies. Here we will use a survey with specifications
similar to those of ACT, but covering 4,000 square degrees and thus
with higher noise. These specifications are not too dissimilar from
those of SPT, assuming that accurate photometric redshifts can be
obtained for all SPT clusters.

\subsection{Probing the kSZ flux at high redshift}

In what follows, we shall use specifications for ACT one year data 
(Gaussian beam with FWHM equal to 2 arcmins, noise amplitude lower than 2 $\mu$K per beam, and a clean scanned area of 400 square degrees).
ACT observes in three bands: 145, 220 and 250 GHz. We will concentrate
on the 220 GHz channel, although some knowledge will be assumed to be
inferred from the other bands. For instance, the 250 GHz channel will
be very useful to estimate the level of infrared-galaxy emission at
lower frequencies. Likewise, the 145 GHz channel should be critical
when characterizing the tSZ flux from each cluster. For simplicity, we
will concentrate on the variance of the kSZ signal, but it is easy to
estimate the kSZ angular correlation function introduced in subsection
4.4. \\

The strip covered by ACT should also be surveyed by
SALT\footnote{SALT's URL site: {\tt http://www.salt.ac.za/}}.  Cluster
detection via tSZ effect will provide targets for optical
observations. Alternatively, optical cluster identification should be
possible up to $z\sim 1$ from SALT's multi-band imaging using
algorithms such as those developed in \citet{kim,c4}. Hence, a direct
comparison can be made with tSZ detected cluster sample.  SALT
spectroscopic follow up will enable to obtain the cluster redshifts. 

%
%

\begin{figure}
\begin{center}
         \epsfxsize=8cm \epsfbox{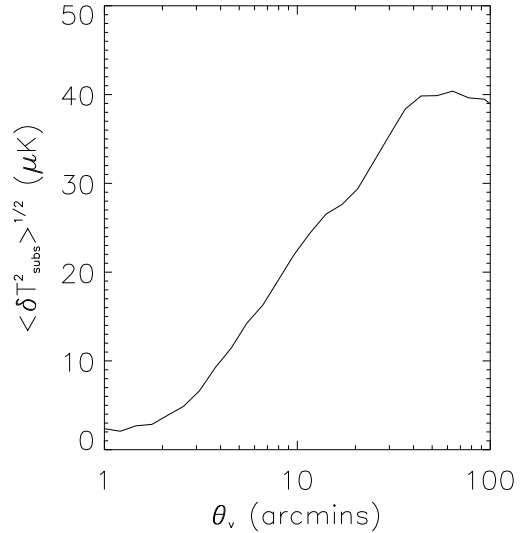}
\caption[fig:cmb_res]{Typical amplitude of CMB residuals remaining
after attempting to approximate the CMB average within a circular
patch of angular radius $\theta_v$ (given in abscissas) by the CMB
average computed within a ring surrounding the patch of width 10\% the
patch radius. For most of clusters located at $z > 0.3$, these
residuals will typically be of 3 -- 4 $\mu$K amplitude.}
\label{fig:cmb_res}
\end{center}
\end{figure}

The method is as follows:

\begin{itemize}

\item For every detected galaxy cluster, we take a patch of radius
equal to one projected cluster virial radius.  We can use the edge of
significant tSZ emission at other frequencies to define this
radius. We compute the mean temperature within this patch, and draw a
ring surrounding it, of width, say, 10\% of the virial radius.  We
next compute the mean temperature within this ring and subtract it
from the mean of the patch. This operation should remove most of the
CMB contribution to the average temperature in the patch, but will
unavoidably leave some residuals, which will be denoted here as
$\delta T_{cmb}^{res}$.  These residuals will have two different
contributions: the first coming from the inaccurate CMB subtraction,
the second one being due to instrumental noise residuals,
\begin{equation}
\langle \bigl( \delta T_{CMB}^{res}\bigr)^2 \rangle = 
  \langle  \delta T_{subs}^2 \rangle + \frac{N^2}{N_{beams}}.
\label{eq:cmb_res1}
\end{equation}
$N_{beams}$ is the number of beam sizes present in the ring, and
$\vnn^2$ is the instrumental noise variance. If we denote the patch of
radius the cluster virial radius as {\it region 1}, and by {\it region
2} the ring surrounding it, it can be easily proved that the first
term in the RHS of the last equation reads
\[
 \langle  \delta T_{subs}^2 \rangle = \frac{1}{(\Delta \Omega_1)^2}
 \int_{\Delta \Omega_1}\int_{\Delta \Omega_1}\; d\vn_1 d\vn_2\; C(\vn_1,\vn_2) \; + \;
\]
\[
\phantom{xxxxxxxx}
\frac{1}{(\Delta \Omega_2)^2} 
         \int_{\Delta \Omega_2}\int_{\Delta \Omega_2}\; d\vn_1 d\vn_2\; C(\vn_1,\vn_2) \; - \;
\]
\begin{equation}
\phantom{xxxxxxxx}
\frac{2}{\Delta \Omega_1 \Delta \Omega_2} 
  \;\;\; \; \int_{\Delta \Omega_1} \int_{\Delta \Omega_2} \; d\vn_1 d\vn_2 \;  C(\vn_1,\vn_2).
\label{eq:cmb_res2}
\end{equation}
$\Delta \Omega_1$ and $\Delta \Omega_2$ denote the solid angles of the
patch and the ring, respectively, whereas $\vn_1$ and $\vn_2$ denote
directions in the sky and $C(\vn_1,\vn_2)$ is the CMB angular
correlation function evaluated at the angle separating the directions
$\vn_1$ and $\vn_2$.  The rms fluctuations introduced by this residual
are plotted in Fig.(\ref{fig:cmb_res}): although it can be as high as
a few tens of microK for nearby clusters subtending 30 -- 40 arcmins,
their effect reduces to a 3 -- 4 $\mu$K for clusters of a few arcmins
size, which correspond to most of clusters at $z > 0.3$ to be detected
by ACT-like CMB experiments. This contaminant will be dominant over
the instrumental noise contribution.\\

\item We assume that, by observing the tSZ amplitudes at 145 GHz and
  250 GHz, it is possible to provide an {\em unbiased} estimate of the
  mean tSZ contribution to the cluster patch at 220 GHz. After
  subtracting the tSZ and CMB components estimates, the temperature in
  the patch corresponding to cluster $i$ can be written as
\begin{equation}
 \delta_i = \delta T_{CMB,i}^{res} + \delta T_{tSZ,i} + \vnn + \delta
 T_{kSZ,i}^{int} + T_{kSZ,i}.
\label{eq:delta_i}
\end{equation}

$\delta T_{tSZ,i}$ denotes the tSZ residuals {\em after} substracting
the estimated tSZ amplitude. $\vnn$ accounts for contribution of
instrumental noise to the patch average, and $\delta T_{kSZ,i}^{int}$
for the residual contribution of internal velocities. According to
\citet{diaferioetal}, we assume a typical rms for internal
velocities of one third the bulk flow velocity expected for each
cluster, and random orientation (sign); thus we assume $\delta
T_{kSZ,i}^{int}$ rms to be $1/3$ of the bulk-flow induced kSZ
amplitude.  The first four quantities in the RHS of eq.(\ref{eq:delta_i}) 
have zero mean, hence $\delta_i$ is an {\em unbiased}
estimate of the kSZ amplitude in cluster $i$.\\

\item We now combine estimates of the kSZ coming from {\em different}
  clusters in order to estimate the kSZ cluster-cluster correlation
  function, $\langle T_{kSZ,i} T_{kSZ,j} \rangle$. As mentioned in a
  previous Section, we are interested in computing this correlation
  function by considering pairs of clusters of similar redshift.  Note
  that we avoid squaring the kSZ estimate of the same cluster ($i=j$),
  since residuals would not average out and would introduce a bias in
  our kSZ variance estimates.  Let us assume that the bulk flows
  occupy a typical scale (named here as {\it coherence scale})
  $\theta_{coh}$ on the sky.  Within this patch along direction $\vn$,
  we can sort all clusters in mass and redshift bins. Let $I$, $J$, $L$ and $M$
  denote the mass bins of clusters in a common redshift range centered
  at $z_l$.  By building cluster pairs upon all members of these bins,
  and combining different mass bins, it is possible to compute the
  following kSZ variance\footnote{We integrate temperature 
  fluctuations in the cluster's area, obtaining a quantity of units
  $\mu$K$\cdot$strad. This integrated temperature fluctuations are essentially
  proportional to {\em flux} fluctuations, and for this reason they will be
  denoted by $F_{kSZ}$. One must keep in mind, however, that their units are
  not Janskys, but microKelvin $\times$ stereoradian.} 
  estimator at redshift $z_l$ along the
  direction $\vn$:
\begin{equation}
 \tilde{F}_{kSZ,\; l}^2 (\vn) \equiv  \sum_{I\leq J}
\frac{w_{I,J}}{N_{I,J}} \sum_{i,j} (\delta_i\Omega_i) (\delta_j\Omega_j)
       \bigg/ \sum_{I\leq J} w_{I,J}.
\label{eq:est1ksz}
\end{equation}
 $\Omega_{j}$ stands for the {\it j-th} cluster's solid angle. 
 We remark that the
 kSZ flux depends solely on the cluster's mass and peculiar velocity, and
 so do our angle-integrated kSZ temperature anisotropies.
 If we denote by $N_I$ and $N_J$ the number of cluster members in bins
 I and J, respectively, then the number of cluster pairs that can be
 formed by combining these two mass bins is given by $N_{I,J}$;
 $N_{I,J}=N_I N_J$ if $I\neq J$ and $N_{I,J}=N_I (N_I-1)/2$ for the
 same bin ($I=J$).  The indexes $i$ and $j$ run for individual cluster
 members in each mass bin.  $w_{I,J}$ is a weight factor, which we
 define as
\[
w_{I,J} \equiv \frac{N_I N_J}{\sigma_I^2\sigma_J^2} = 
\]
\[
 \phantom{xxx} \frac{N_I}
      {\Omega_I^2[\langle \delta T_{CMB}^{res}\rangle^2 +
 \langle \delta T_{tSZ} \rangle^2 +  \langle (\delta T_{kSZ}^{int})^2 \rangle 
                             + \vnn^2]_I} \;\; \times
\]
\begin{equation}
  \phantom{xxx}
 \frac{N_J}{
 \Omega_J^2[\langle 
     (\delta T_{CMB}^{res})^2 \rangle + \langle \delta T_{tSZ}^2 \rangle + 
                    \langle (\delta T_{kSZ}^{int})^2 \rangle +     \vnn^2]_J},
\label{eq:wij}
\end{equation}
 with the subscripts $I$ and $J$ evaluating the brackets in the
 corresponding mass bins.  The estimator of eq.(\ref{eq:est1ksz})
 provides a weighted measurement of the kSZ variance,
\begin{equation}
\langle   \tilde{F}_{kSZ,\; l}^2 (\vn) \rangle = 
\; \frac{\sum_{I\leq J}w_{I,J}\; (F_{kSZ,\;l,I}F_{kSZ,\;l,J})}{\sum_{I\leq J}w_{I,J}},
\label{eq:avest1ksz}
\end{equation}
 with $F_{kSZ,\;l,I}$ the expected kSZ amplitude at redshift $z_l$ and
mass equal to that corresponding to bin $I$.  Its formal error is
given by
\[
\Delta^2 [\tilde{F}_{kSZ,\; l}^2(\vn)] = 
\]
\[
\phantom{xxxx} 2\times \; \biggl[ \frac{\sum_{I\leq J}w_{I,J}\;
(F_{kSZ,\;l,I}F_{kSZ,\;l,J})}{\sum_{I\leq J}w_{I,J}}\biggr]^2 \;\; +
\]
\[
\phantom{xxxxxxxxxx}
\sum_{I,J,L} w_{I,J,L} (F_{kSZ,\;l,I}F_{kSZ,\;l,L})
  \bigg/ \biggl( \sum_{I\leq J} w_{I,J}\biggr)^2 \; +
\]
\[
\phantom{xxxxxxxx}
\sum_{I,J,M} w_{I,J,M} (F_{kSZ,\;l,J}F_{kSZ,\;l,M})
  \bigg/ \biggl( \sum_{I\leq J} w_{I,J}\biggr)^2 \; +
\]
\begin{equation}
\phantom{xxxxxxxxxx}
  1 \bigg/ \sum_{I\leq J} w_{I,J},
\label{eq:errest1ksz}
\end{equation} 
where $w_{I,J,L} \equiv w_{I,J} N_L / \sigma_L^2$.  Note that the
first term in the RHS of this equation is not sensitive to the number
of clusters within the coherence patch. Such term, containing the
squared kSZ expectations for mass bins $I$ and $J$, is associated to
the (assumed) intrinsic Gaussian nature of the kSZ fluctuations, and
it corresponds to the {\it cosmic variance} contribution. It will thus
scale like the inverse of the survey area when different coherence
patches are combined in the analysis.  The forth term is exclusively
due to observational errors, whereas the second and third terms are
hybrid: they show contributions from both the intrinsic uncertainty of
the velocity field and observational errors.  If $N_b$ denotes the
number of mass bins, and we take the weights, the cluster number and
the $F_{kSZ}$ equal for all mass bins ($\sigma_I = \sigma$, $N_I = N$
and $F_{kSZ,l,I} = F_{kSZ,l}$ for every mass bin $I$), then it can
easily be proved that the forth term scales roughly as $\sigma^4 / N^2
/ (N_b(N_b+1))$, that is, the squared variance expected for each
cluster over the total number of pairs that can be formed. The second
and third terms, in this limit, yield $\sigma^2 F_{kSZ,l}^2
/(N(N_b+1))$, which scales inversely to the number of clusters. In our
case, our kSZ flux estimates will be limited by the cosmic variance
term. On the other hand, because of having very few objects, we
considered one mass bin only.\\

\item Finally, we combine estimates from the $N_{coh} \simeq 4 \pi
f_{sky} / \theta_{coh}^2$ different projected coherent regions in the
sky, where $f_{sky}$ is the fraction of the sky covered by the CMB
experiment, yielding
\begin{equation}
 \tilde{F}_{kSZ,\; l}^2 \equiv \sum_{\vn}\frac{\tilde{F}_{kSZ,\;
 l}^2(\vn)}{\Delta^2 [\tilde{F}_{kSZ,\; l}^2(\vn)]} \bigg/ \sum_{\vn}
 \frac{1}{\Delta^2 [\tilde{F}_{kSZ,\; l}^2(\vn)]},
\label{eq:est2ksz}
\end{equation}
with an uncertainty 
\begin{equation}
\Delta^2 [\tilde{F}_{kSZ,\; l}^2] = \;\;\; 1 \bigg/
  \sum_{\vn} \frac{1}{\Delta^2 [\tilde{F}_{kSZ,\; l}^2(\vn)]}.
\label{eq:errest2ksz}
\end{equation} 
Note that the size of the coherent patch must depend of redshift: a bulk
flow extenting up to 20 $h^{-1}$Mpc at z=1 subtends a degree on the
sky, whereas if it is at z$\simeq 0.05$ then it subtends around 8 degrees.

\end{itemize}

We take the effective noise $\vnn$ to have a typical amplitude of 5
$\mu$K per beam\footnote{The noise level for ACT is expected to be $
  ^{<}_{\sim} 2 \mu$K per beam.}, and it accounts for both
instrumental noise and the confusion noise associated to unresolved
sources.  $\delta T_{CMB}^{res}$ is computed as stated in
eq.(\ref{eq:cmb_res2}), and contributes typically with a few microK.
Regarding the tSZ residuals, $\delta T_{tSZ}$ contains the
contribution coming from relativistic tSZ corrections, and power
leakage associated to the finite spectral width of the detectors. We
conservatively assume that the amplitude of these residuals (remaining
after the tSZ substraction) is typically the non-relativistic tSZ
temperature increment expected at 222 GHz.  We approximate the beam as
a Gaussian of FWHM equal to 2 arcmins, and take 400 square degrees
($f_{sky} \simeq 10^{-2}$) as the clean sky region covered by ACT.
ACT's sensitivity limit is set to cluster above
$2\times 10^{14}$ $h^{-1}M_{\odot}$, and the total
number of clusters above this threshold predicted in this region of
the sky by our Sheth-Tormen mass function is roughly 4,400. However,
for this analysis we only use relatively bright and big clusters: the
product of their angular size and tSZ temperature decrement at 145 GHz
must be bigger than 160 $\mu$K arcmin$^2$, and this requirement
decreases considerably
the amount of available clusters.\\

\begin{figure}
\begin{center}
         \epsfxsize=8cm \epsfbox{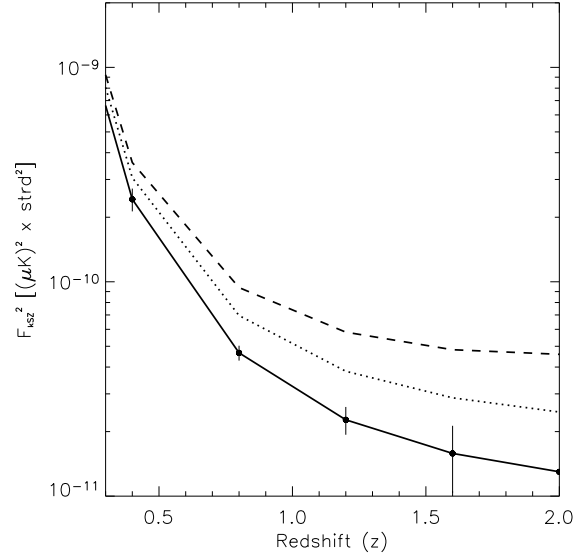}
\caption[fig:errrel]{Expected amplitude of the squared kSZ flux for
ACT, after grouping all cluster in six redshift bins. The solid line
line corresponds to a $\Lambda$CDM. The dotted and the dashed lines
display the kSZ variances for universes with Dark Energy components
having $w$ equal to -0.6 and -1/3, respectively}
\label{fig:errrel}
\end{center}
\end{figure}

As a result of this section, in Fig.(\ref{fig:errrel}) we show our
expectations of ACT's sensitivity on the kSZ  variance
when all clusters are grouped in the redshift bins
 $z^{band}_{l} \equiv $ [0.02, 0.4, 0.8,
1.2, 1.6, 2.0]. Solid, dotted and dashed lines
correspond to a $\Lambda$CDM model($\Omega_m = 0.3,
\Omega_{\Lambda}=0.7$), a flat universe with dark energy equation of state 
$w=-0.6$, and $w=-1/3$, respectively. We report error bars for the $\Lambda$CDM
model: we drop the first point at $z=0.02$ (which is dominated by cosmic
variance) and focus in the high redshift range: the signal-to-noise ratio
shows a maximum at z=0.8, and beyond this redshift the error bars start to
increase due to the lack of massive clusters.

\subsection{The kSZ/tSZ ratio ${\cal R}$}

In this subsection, we address the study of the ratio of the kSZ and
tSZ effects. When referring to temperature anisotropies, the kSZ and
tSZ effects are integrals weighted by the cluster electron density
along the cluster diameter, and are sensitive to the size and/or the
concentration parameter of these objects. The ratio of the kSZ and the
tSZ, however, should cancel these dependencies out to great extent, and
provide a cleaner view of the cluster's temperature
and peculiar velocity.\\

Following our cluster model, we investigate the behaviour of the statistic
${\cal R} \equiv \delta T_{kSZ} / \delta T_{tSZ}$ in the cluster population.
If the evolution of the cluster temperature is well described by the spherical
collapse model, we shall find then that ${\cal R}$ should be a good estimator
of the dark energy equation of state $w$ at recent epochs. But since cosmic
variance becomes important at low redshifts (smaller volume  for a given
solid angle), in this case we shall compute
our expectations for an  experiment  with sky coverage close to
$f_{sky} = 0.1$ (4,000 square degrees). To compensate for the wider
survey area we assume a noise of 10 $\mu$K per arcminute squared and
that clusters up to $z= 0.8$ can be detected and  resolved. We
assume that the  frequency coverage enables contaminant subtraction  and
extraction of clusters tSZ signal. We also assume that follow up
observations will yield redshift for all observed clusters. These
specifications are not too dissimilar to those of the SPT telescope
when combined with photometric follow up.

We shall see that these are precisely the requirements needed to
obtain cosmological information from ${\cal R}$. However, since in
this case the kSZ signal is divided by the tSZ temperature decrement
(measured at say, 145 GHz), we must be very careful with the noise
contribution to the denominator of ${\cal R}$. Our approach will be to
conservatively consider only clusters whose integrated tSZ temperature
decrements are larger than 160 $\mu$K-arcmin$^2$ at 145 GHz.  This
implies that tSZ errors will be typically 5\% -- 10\% of the estimated
tSZ temperature decrement, and that they can be treated perturbatively
as errors in the kSZ estimation in the numerator.  Therefore, our
model for the estimation of ${\cal R}$ in a single cluster will be
given by:

\begin{equation}
 r_i = \frac{\delta T_{CMB,i}^{res} + \delta T_{tSZ,i} + \vnn + \delta
 T_{kSZ,i}^{int} + \epsilon\;T_{kSZ,i} + T_{kSZ,i}}{T_{tSZ,i}}.
\label{eq: r_i}
\end{equation}

Most of the terms of this equation are defined exactly as 
in eq.(\ref{eq:delta_i}). The only new term is $\epsilon T_{kSZ,i}$, 
which accounts
for the extra error introduced by the uncertainty in the denominator,
$T_{tSZ,i}$. Here $T_{tSZ,i}$ is the absolute value of the tSZ decrement of
the cluster at 145 GHz  and $\epsilon$ was taken to be a normal random variable
of rms 0.05 (5\% error in $T_{tSZ,i}$ which reflects into 5\% error in 
$T_{kSZ,i}$). We neglect the correlation of the errors in
$T_{tSZ,i}$ with those of $T_{kSZ,i}$ and take $\epsilon$ independent of 
the other noise sources.

We now write the analogue to eq.(\ref{eq:est1ksz}) as
\begin{equation}
 \tilde{\cal R}_{l} (\vn) \equiv  \sum_{I\leq J}
\frac{w_{I,J}}{N_{I,J}} \sum_{i,j} r_i r_j
       \bigg/ \sum_{I\leq J} w_{I,J},.
\label{eq:est1rat}
\end{equation}

where the weights are defined as
\[
w_{I,J} \equiv \frac{N_I N_J}{\sigma_I^2\sigma_J^2} = 
\]
\[
 \phantom{} \frac{ N_I\;T_{tSZ,I}^2}
      {[\langle \delta T_{CMB}^{res}\rangle^2 +
 \langle \delta T_{tSZ} \rangle^2 +  \langle (\delta T_{kSZ}^{int})^2 \rangle 
             +  \epsilon^2 T_{kSZ,I}^2  + \vnn^2]_I} \;\; \times
\]
\begin{equation}
  \phantom{}
 \frac{N_J\;T_{tSZ,J}^2}{
 [\langle 
     (\delta T_{CMB}^{res})^2 \rangle + \langle \delta T_{tSZ}^2 \rangle + 
                    \langle (\delta T_{kSZ}^{int})^2 \rangle + 
                    + \epsilon^2 T_{kSZ,J}^2 + \vnn^2]_J}.
\label{eq:wij_r}
\end{equation}

The estimate of ${\tilde {\cal R}}_l (\vn)$ in a redshift band $z_l$
and direction $\vn$, together with its uncertainty $\Delta^2 [{\tilde
{\cal R}}_l (\vn)]$ can be obtained from
eqs.(\ref{eq:avest1ksz},\ref{eq:errest1ksz}) by simply replacing
${\tilde F}_{kSZ,l}^2(\vn)$ by ${\tilde {\cal R}_l (\vn)}$.  Similarly
the expressions for ${\tilde {\cal R}}_l$ and $\Delta^2 [{\tilde {\cal
R}}_l]$ can be obtained from
eqs.(\ref{eq:est2ksz},\ref{eq:errest2ksz}).\\

\begin{figure}
\begin{center}
         \epsfxsize=8cm \epsfbox{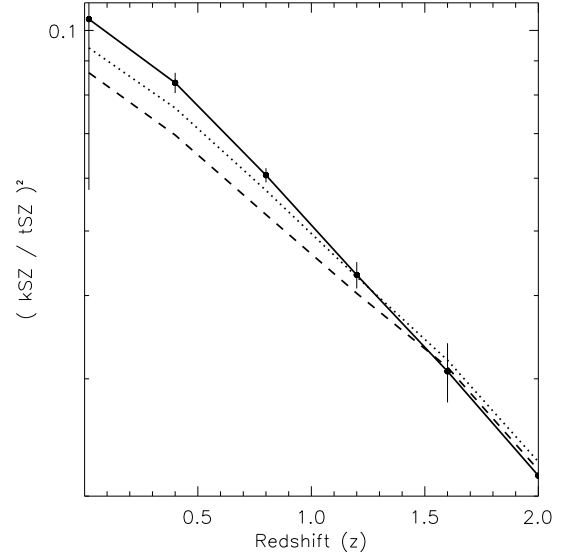}
\caption[fig:ratio_mod]{Expected kSZ/tSZ ratio ${\cal R}$ as measured by
SPT for six redshift bins. The solid line corresponds to a
$\Lambda$CDM model while the dotted and the dashed lines are for flat
models with  $w$ equal to -0.6 and -1/3, respectively.}
\label{fig:ratio_mod}
\end{center}
\end{figure}

In Fig.(\ref{fig:ratio_mod}) we plot the ratio ${\cal R}$ at different
redshifts as it would be seen by an experiment like SPT. As before, we have 
assumed that only clusters above $2\times 10^{14}$ $h^{-1} M_{\odot}$ can
be seen, and grouped all clusters in the redshift bins  $z^{band}_{l} \equiv $ [0.02, 0.4, 0.8, 1.2, 1.6, 2.0]. The lines refer to the same cosmological 
models as in Fig.(\ref{fig:errrel}). Note the similarity of this plot and
Fig.(\ref{fig:Dv}).  We see that ${\cal R}$ is sensitive to cosmology at 
much lower redshifts ($z ^<_{\sim}\;0.8$) than $F_{kSZ}$, and for the
sensitivity analysis following in the next Section we have only used the
first three redshift bins.

\section{kSZ Sensitivity to Measuring Cosmological Parameters}

\begin{figure}
\begin{center}
         \epsfxsize=8cm \epsfbox{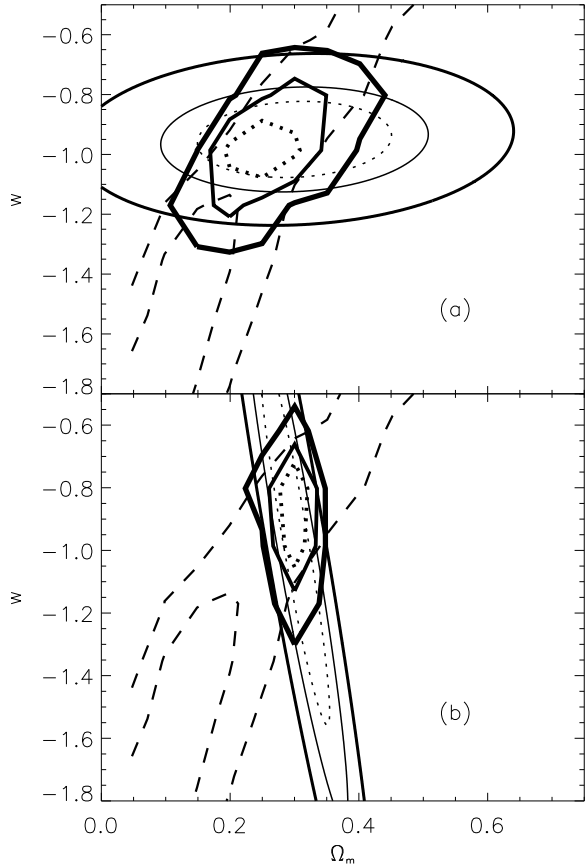}
\caption[fig:like_con]{Contour plots showing the 1 and 2--$\sigma$
  joint confidence region (solid line) and the 1--$\sigma$
  marginalized (dotted line)  in the $\Omega_m$ -- $w$ plane for the
  $F_{kSZ}^2$ (method (a)) and ${\cal R}^2$
 (method (b)) estimations. Method (a) applied to ACT (top)  with CMB (WMAP 1st
  year)  priors (dashed lines) yields a $\sim 8$\% error on $w$. This error
 would drop below 5\% if ACT covers 1,000 square degrees.
  Method (b) on a SPT-like survey gives a $\sim 12$\% error on $w$ after imposing the prior from WMAP's 1st year data.} 
\label{fig:like_con}
\end{center}
\end{figure}

We  can now explore the dependence of the two statistics introduced in the
previous Section on cosmological parameters. 
We define the $\chi^2$ as 
\begin{equation}
\chi^2 \equiv \sum_l
      \frac{ \biggl( \tilde{Q}_{kSZ,\; l}^2  
                  - Q_{kSZ,\; l}^2 \biggr)^2 } 
         { \Delta^2 [Q_{kSZ,\; l}^2] }, 
\label{eq:chisq}
\end{equation}

where $ Q_{kSZ,\; l}^2$ can either refer to the angle-averaged kSZ
anisotropy ($F$) or the kSZ/tSZ ratio (${\cal R}$).  The likelihood is
thus ${\cal L}\propto \exp -\frac{1}{2} \chi^2$.  We estimate errors
using the Fisher matrix approach.  In principle the parameters that
enter in the analysis are $\Omega_m$, $w$, the fraction of cluster
mass in the intra cluster medium $f_{ICM}$, the present-day
normalization of the matter density fluctuations $\sigma_8$ and the
reduce Hubble constant $h$.  In practice both methods are non sensitive
to $f_{ICM}$, $\sigma_8$ and $h$ separately for a fixed number of
detected clusters,
but on their combination in the form of an overall amplitude of the
kSZ signal, which we shall asign to an amplitude parameter $A$ alone. 

We assume that $A$ is redshift-independent (or that
its scaling in redshift can be constrained) and consider a 20\%
uncertainty in this normalization, owing to 10\% uncertainty in $\sigma_8$,
$f_{ICM}$ and $h$ each, over which we marginalize. We remark, however, that
our results will be practically insensitive to the uncertainty on $A$.  
In Fig.(\ref{fig:like_con}) we show the resulting constraints in the
$\Omega_m$--$w$ plane. Clearly, the $F_{kSZ}$ method is more sensitive to
$w$ than ${\cal R}$, and their directions of degeneracy are also different,
but both distinct to the directions of degeneracy corresponding to estimators
based upon Large Scale Structure. Indeed, these estimators are restricted to
the low redshift universe, and hence their sensitivity on $w$ is very
limited, giving rise to degeneracy direction almost parallel to the $w$ axis,
(see, e.g., Fig~(11) in \citet{eisenstein} or Fig.~(13) in \citet{tegmark04}).
The constraint on $w$, marginalized over
$\Omega_m$, is $\sim$ 12\% for $F_{kSZ}$ and $\sim$ 60\% for ${\cal R}$. As the
degeneracy direction in each case is different to that corresponding to 
CMB temperature measurements,
 a combination with analyses of CMB temperature data can break the degeneracy.
When considering WMAP first year data, the marginalised error on $w$ reduces to
$\sim 8$\%, $\sim 12$\%, for $F_{kSZ}$ and ${\cal R}$ methods, respectively 
(thick dotted lines). As pointed above, these errors are dominated by the
cosmic variance term present in eq.(\ref{eq:errest1ksz}). Therefore, 
by increasing the sky covered by the experiments, those errors
 should decrease as $\propto 1/\sqrt{f_{sky}}$.\\

 These results have been obtained after using a ST mass function,
  and assuming that the cluster density was uniform (i.e., we have
  have considered no biases). The error amplitudes and the
  orientation of the ellipses in Fig (\ref{fig:like_con}) depend
  strongly on the number of cluster pairs that can be formed in each
  redshift bin, particularly at the high z end. If the actual
  sensitivity of future CMB experiments is such that the lower limit
  on detectable clusters can be relaxed (we believe the limit $2\times
  10^{14}\;h^{-1}$ M$_{\odot}$ to be very conservative for ACT's noise
  level), then the resulting number of cluster pairs that can be
  formed would increase considerably, and this would reflect on an increased
  sensitivity to $w$. \\

\begin{figure}
\begin{center}
         \epsfxsize=8cm \epsfbox{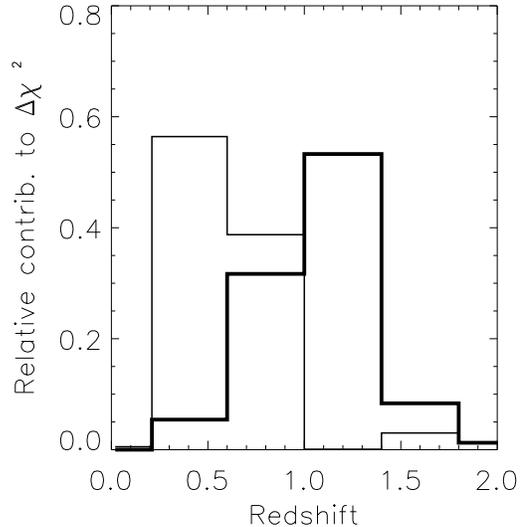}
\caption[fig:like_vs_z]{Differential redshift contribution to
the  change in $\chi^2$ when probing models with different $w$:
approach a) is more sensitive to higher redshifts (thick line),
method b) to lower redshifts (think line).}
\label{fig:lik_vs_z}
\end{center}
\end{figure}

We note that, a priori, the two methods are nicely complementary in their
redshift ranges sensitive to $w$ as shown in
Fig.(\ref{fig:lik_vs_z}): we find that the $F_{kSZ}^2$ estimation 
(method a) thick histogram) is sensitive to $w$ at high redshifts ($z>
0.5$), whereas ${\cal R}$ for different dark energy models
differs at low redshift and converges at  $z \approx 1$: the
sensitivity of this method is localised mainly at low redshift (thin
histogram). \\

\begin{figure}
\begin{center}
         \epsfxsize=8cm \epsfbox{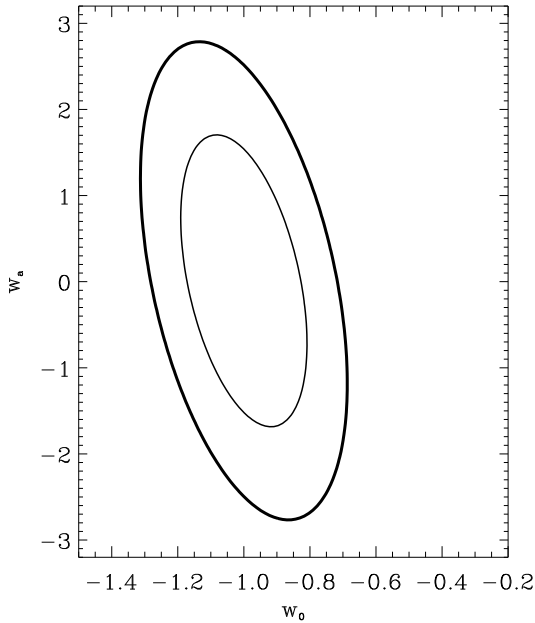}
\caption[fig:w_a]{Marginalisation in the $w_0$-$w_a$ plane for the ACT-like
experiment. $w_a$ expresses the variation of $w$, which is parametrised as
$w(a) = w_0 + w_a(1-a)$, with $a$ the scale factor. Contours correspond to
1 \& 2-$\sigma$ joint confidence regions.}
\label{fig:w_a}
\end{center}
\end{figure}

After parametrising the evolution of $w$ as 
$w(a) = w_0 + w_a(1-a)$, with $a=1/(1+z)$ the scale factor, we conduct
a Fisher matrix analysis considering the
parameter set [$\Omega_m$, $w_0$, $f_{ICM}$, $w_a$] for the $F_{kSZ}^2$ method.
The marginalization in the $w_0$-$w_a$ plane is given in Fig.(\ref{fig:w_a}),
which shows this method should have some residual sensitivity on $w_a$ (a
typical error of $\sim 1.75$ in $w_a$). Note that this error should be 
improved further after combining the two methods proposed here, provided that 
they are sensitive to different redshift ranges. 

\section{Conclusions}

We have studied bulk flows in the large scale structure in the context
of the kSZ effect and future high-sensitivity and high-resolution CMB
experiments like ACT.  Since the kSZ effect is only sensitive to the
projected peculiar velocities of electron clouds, we have focussed our
analysis on the angular correlation of  radial peculiar
velocities in galaxy clusters. We  have provided an analytical expression for
the angular correlation function of projected peculiar velocities in
the linear regime, and
interpreted it in the context of local and distant bulk flows. 
 We have also presented an expression for the power spectrum of
 projected linear peculiar velocities.\\
 
 We have investigated in which redshift and for which cluster mass
 ranges large-scale bulk flows should be more easily detectable in
 future kSZ surveys, and computed the overall effect of the entire
 galaxy cluster population on the CMB sky. We have shown that the main
 contribution comes from Poissonian/random fluctuations of the number
 of clusters along the line of sight, especially at small angular
 scales. However, we find that the local bulk flow generates a signal
 which should be dominant at the very large angular scales of the
 quadrupole and octupole. Other terms associated to the coupling of
 velocity with density fluctuations  give smaller contributions. 
We have calculated the kSZ signal for the cluster
sample accessible by  forthcoming experiments, and considered
different sources of contamination which may limit our capacity to
distinguish the kSZ signal from other components. \\

We have presented two approaches to measure the kSZ signal and exploit
its dependence on cosmological parameters such as the equation of
state of dark energy ($w$). The first method is based on measurements of
the kSZ flux and its redshift evolution. Its sensitivity to cosmology
increases with redshift. For this reason it is suited for ``deep and
narrow'' survey strategies. A data set such as that provided by ACT 2
year observations with SALT follow up is well suited for an application
of this method. ACT can detect the kSZ signal with a S/N $\sim 12$ at
$z\sim 1$.\\

The second approach uses the  ratio of kSZ and tSZ-induced
temperature anisotropies. The cosmology-dependence  is strongest at
low redshift and hence this method is suitable for ``wide and shallow''
survey strategies. In this case we have considered an ACT-like experiment with
larger sky coverage ($1/10$ of the sky) and increased instrumental
noise. These are specifications similar to those of SPT with at least
three frequency bands and combined with redshift determinations of the detected
clusters. In this case the kSZ effect can be detected with a S/N of
$\sim$ 30.\\

These methods can yield  constraints on cosmological parameters, in
particular can constrain  the equation of state for dark energy  at
the 10\%  level. The two methods are nicely complementary as they
measure dark energy in different redshift ranges, opening up the
possibility to constrain dark energy redshift evolution.

\section*{Acknowledgments}

We are indebited to Ravi Sheth for help and enlightening discussions
on cluster spatial correlations and non linear velocities. LV thanks
Mark Devlin for discussions.  We thank Simon Dedeo for comments and
discussions.  This research is supported in part by grant NSF
AST-0408698 to the Atacama Cosmology Telescope.  CHM and LV are
supported by NASA grants ADP03-0000-0092 and ADP04-0000-0093. The work
of RJ is supported by NSF grants AST0206031, AST-0408698, PIRE-0507768
and NASA grant NNG05GG01G.  DNS is supported by NSF grants
PIRE-0507768 and through the NASA ATP programs and the WMAP project.

\appendix

\section{Appendix A}

\vspace{.2cm}

\noindent {\bf Coupling the linear density and velocity fields}
\vspace{.4cm}

In linear theory, the density contrast $\delta(\vx) \equiv (\rho(\vx) 
- \bar{\rho}) / \bar{\rho} $ is still much smaller than unity. This
allows linearising the evolution equations and neglecting all non-linear
orders, which makes the the evolution of each
Fourier mode $\delta_{\vk}$ is independent from the other modes.

In a Friedmann Robertson Walker (FRW) universe, the perturbed continuity
equation reads
\begin{equation}
\frac{\partial \delta (\vx)}{\partial t} + \frac{\nabla \vv (\vx)}{a} = 0,
\label{eq:cont}
\end{equation}
which can be re-written in Fourier domain as
\begin{equation}
\frac{\partial \delta_{\vk}}{\partial t} = -\frac{i}{a} \vk\cdot\vv_{\vk},
\label{eq:cont2}
\end{equation}
 with $a=1/(1+z)$ the scale factor.
Throught this paper, the velocities are {\em proper} peculiar velocities.
We must note that the expansion of the velocities in terms of 
its Fourier modes is given by 
\begin{equation}
 \vv (\vx ) = \int \frac{d\vk}{(2\pi)^3} \; \vv_{\vk} \;\exp{-i\vk\vx},
\label{eq:vk}
\end{equation}
and that, due to isotropy, the diferent components are independent. The
statistical properties of each of them must be, however, identical.

Coming back to eq.(\ref{eq:cont2}), 
if we denote by ${\cal D}_{\delta}(z)$ the growth factor of the density
perturbations, then we can express the
component of $\vv_{\vk}$ parallel to $\vk$ (denoted here by $v_{\vk}$) as
\begin{equation}
v_{\vk} = i\; H(z) \frac{d{\cal D}_{\delta}}{dz} \;\frac{\delta_{\vk}}{k}.
\label{eq:vkmode}
\end{equation}
$H$ stands for the Hubble function and $z$ for redshift.

The power spectrum for {\em any} component of the Fourier velocity mode
hence reads
\begin{equation}
P_{vv}(k) \; = \;\biggl( H(z)\frac{d{\cal D}_{\delta} (z)}{dz} \biggr)^2
 \frac{P_{m}(k)}{k^2}.
\label{eq:vvps}
\end{equation}
If we now combine $\delta_{\vk}$ and $v_{\vq}$, we obtain
\begin{equation}
\langle \delta_{\vk} v_{\vq}^* \rangle = (2\pi)^3 \delta^D(\vk-\vq)
P_{dv}(k)  = 
(2\pi)^3 \delta^D(\vk-\vq) \;
i\; H(z)\frac{d{\cal D}_{\delta} (z)}{dz} \frac{P_{m}(k)}{k}.
\label{eq:dvps}
\end{equation}
Note that this is an imaginary quantity. In this work, we shall find this
power spectrum either squared or 
in a convolution with itself, giving rise to negative power 
and indicating {\em anticorrelation}.\\

Finally, we compute the power spectrum of the quantity
\begin{equation}
\phi (\vx, \vn_1) \equiv \delta(\vx)\; (\vv(\vx)\cdot \vn_1)
 = \sum_i \delta (\vx) 
      \bigg( v_i(\vx) n_i^1\biggr),
\label{eq:phi}
\end{equation}
where $\vn_1$ stands for the unitary vector connecting an observer with
the position $\vx$, and the sum is over the three spatial
components.  Since
a product in real space involves a convolution in Fourier space,
the Fourier mode of $\phi$ reads:
\begin{equation}
\phi_{\vk,\vn_1} = \sum_i \int \dthreeq \;\delta_{\vq}\; v^1_{i,\;\vk-\vq}\;n_i^1.
\label{eq:phiq}
\end{equation}

Having this in mind,
the average product of two Fourier modes of $\phi$ when looking in two
different directions $\vn_1$ and $\vn_2$ is:
\begin{equation}
\langle \phi_{\vk,\;\vn_1} \phi_{\vq,\;\vn_2}^* \rangle = 
    (2\pi)^3 \delta^D(\vk-\vq) \;\; \biggl[
\cos \theta_{12}\bigl[P_{\delta\delta} \star P_{vv}\bigr](k) + 
 \bigl( \cos \theta_{12} + \sin \theta_{12} \bigr) 
\bigl[P_{\delta v} \star P_{v \delta}\bigr](k) \biggr]\;\; +\;
(2\pi)^3 \delta^D(\vk) (2\pi)^3 \delta^D (\vq) 
    \biggl( \int \dthreeu\; P_{\delta v}(u) \biggr)^2 .
\label{eq:phiphips}
\end{equation}
$\theta_{12} = \arccos{(\vn_1\cdot\vn_2)}$ is the angle separating 
$\vn_1$ and $\vn_2$. Without introducing any loss of generality of 
eq.(\ref{eq:phiphips}), we have taken  $\vn_1 = (0,0,1)$ and 
$\vn_2 = (0,\sin{\theta_{12}},\cos{\theta_{12}})$. The symbol $\star$
denotes convolution and is present in the first two terms in brackets.
The third term is non zero only when $\vk=\vq=0$.\\

Before ending this appendix, it is worth to make some remarks upon
the redshift and mass dependence of the power spectra we have computed.
The redshift dependence is explicit via the growth factors ${\cal D}_{\delta}$
and ${\cal D}_{v}$. Now,
if we are interested in the peculiar velocity of a given cluster, then
the (linear) 
velocity field must be averaged in a sphere of dimensions corresponding
to the cluster mass. Something similar can be said about the density field:
when studying the dependence of the number of haloes upon the environment
density, the field $\delta$ must be smoothed on scales which a priori
are dependent on the mass of the haloes whose number density we are studying.
In this scenario, all Fourier modes of the density and the velocity should be
multiplied by the window functions corresponding to the scales within which
we are averaging. This introduces a dependence on 
the masses of the clusters under study in 
$P_{\delta\delta}$, $P_{vv}$ and $P_{\delta v}$.

\vspace{1.cm}

\section{Appendix B}
\vspace{.4cm}

\noindent {\bf The Angular correlation function of the projected velocities}
\vspace{.4cm}

Let $i$ and $j$ be two components of the Fourier mode of the peculiar
velocity field, so that
\begin{equation}
\biggl< v_{\vk}^i (v_{\vq}^j)^* \biggr> = (2\pi)^3 \; \delta^D (\vk-\vq)
   \;\delta_{ij}^K\; P_{vv}(k).
\label{eq:croscomp}
\end{equation}
Using this, we can write the average product of projected velocities as:
\[
\biggl< \biggl(\vv (\vx_1) \cdot \vn_1 \biggr) 
            \biggl( \vv (\vx_2) \cdot \vn_2\biggr) \biggr> = 
\]
\[
\int \; \frac{k^2dk}{(2\pi)^3} \; P_{vv} (k) \;  \biggl(W(kR_1)
W^*(kR_2)\biggr)\int d\phi d(\cos{\theta })
  \cos{\theta_{12}} \; \times 
\]
\begin{equation}
\exp{ -i\;k[x_1 cos\theta - 
          x_2(\sin{\theta}\cos{\phi}\sin{\theta_{12}} + 
                  \cos{\theta_{12}\cos{\theta}})]}.
\label{eq:corv2}
\end{equation}
 
In this equation, the polar axis for the $\vk$ integration has been taken
along the direction given by $\vn_1$. $\theta, \phi$ are the polar
and azymuthal angles of $\vk$, and $\theta_{12}$ is the angle separating
the two directions of observations, $\vn_1$ and $\vn_2$. If we now make use
 of the Rayleigh expansion of the plane wave, i.e.,
\begin{equation}
\exp{-i \vk \cdot \vx} = \sum_l \; (-i)^l (2l+1)\; j_l(kx)\; P_l (\mu),
\label{eq:rayexp}
\end{equation}
(where $\mu$ is the cosine of the angle between $\vk$ and $\vx$ and $P_l$
are Legendre polynomials), and also
the theorem of addition of Legendre functions 
(see, e.g., Gradshteyn \& Ryzhik 8.794), then one ends up with

\begin{equation}
\biggl< \biggl(\vv (\vx_1) \cdot \vn_1 \biggr) 
            \biggl( \vv (\vx_2) \cdot \vn_2\biggr) \biggr> = 
\sum_{even\; l} \frac{2l+1}{4\pi} \; \cos{\theta_{12}} \;\; 
          \biggl( \frac{2}{\pi}\; {\cal F}_l \biggr) \int k^2dk\; 
P_{vv} (k) \;
           W(kR_1)\; W(kR_2) \;  
j_l (k[x_1 - x_2\cos{\theta_{12}}]) 
           \; j_l (k x_2 \sin{\theta_{12}}),
\label{eq:corv2_texb}
\end{equation}
where the factor ${\cal F}_l$ is given by 
\begin{equation}
{\cal F}_l \equiv \frac{(l-1)!!}{2^{l/2} \; (l/2)!} \;\; 
\cos{l\frac{\pi}{2}},
\end{equation}
$j_l(x)$ are the spherical Bessel functions and the summation must take place
only over {\em even} values of $l$. Note that in the limit of 
$\theta_{12} \rightarrow 0$ and $x_1 \rightarrow x_2$, 
this expression becomes eq. (\ref{eq:vel1}).

At the same time, we can rewrite eq.(\ref{eq:corv2}) as an expansion on
Legendre polynomia. Indeed, if we denote by $\vx_1$ and $\vx_2$ the position 
vectors of the clusters, we can use the Rayleigh expansion of the plane 
wave for $\exp{(i\vk\cdot \vx_1)}$ and $\exp{(-i\vk\cdot \vx_2)}$, and write

\begin{equation}
\biggl< \biggl(\vv (\vx_1) \cdot \vn_1 \biggr) 
            \biggl( \vv (\vx_2) \cdot \vn_2\biggr) \biggr> = 
\sum_{l,l'} (2l+1)(2l'+1) (-i)^{l-l'}\int \frac{d\vk}{(2\pi)^3} 
         P_{vv} (k) \; \ctonetwo \;\;
j_l(k x_1) j_l(k x_2) P_l(\mu_{\vk,\vx_1}) 
    P_{l'}(\mu_{\vk,\vx_2}),
\label{eq:corv3}
\end{equation}

with $\mu_{\vx,\vy}$ the cosine of the angle formed by vectors 
$\vx$ and $\vy$. Note also that $\mu_{\vx_1,\vx_2} = \mu_{\vn_1,\vn_2}
= \ctonetwo$. 
Next we apply the addition theorem of Legendre functions on a spherical
triangle formed by $\vn_1$,$\vn_2$ 
and $\hat{\vk}$. As before, we take the polar axis of $\vk$ to be aligned
with $\vx_1$:
\[
\biggl< \biggl(\vv (\vx_1) \cdot \vn_1 \biggr) 
            \biggl( \vv (\vx_2) \cdot \vn_2\biggr) \biggr> =
\sum_{l,l'} (2l+1)(2l'+1) (-i)^{l-l'}\int \frac{d\vk}{(2\pi)^3} 
         P_{vv}(k) \; \ctonetwo\;\;
 j_l(k x_1) j_l(k x_2) P_l(\mu_{\vk,\vx_1}) 
    \biggl[ P_{l'}(\mu_{\vk,\vx_1}) P_{l'}(\mu_{\vx_1,\vx_2})\;\; +
\]
\begin{equation}
\phantom{xxxxxxxxxxxxxxxxxxxxxxxxx} 2\sum_{m=1}^{l'} P_{l'}^m (\mu_{\vk,\vx_1})
  P_{l'}^m(\mu_{\vx_1,\vx_2})\; \cos{(m[\phi_1-\phi_2])} \biggr].
\label{eq:corv4}
\end{equation}

Because $d\vk = k^2dk\sin{\theta}d\theta d\phi_1$, the integration in the
azymuthal angle cancels the sum over $m$ in the brackets. Finally, the
product $\mu_{\vx_1,\vx_2} P_l(\mu_{\vx_1,\vx_2})$ can be rewritten, via a
Legendre recurrence relation, as
a linear combination of $P_{l-1}(\mu_{\vx_1,\vx_2})$ and 
$P_{l+1}(\mu_{\vx_1,\vx_2})$. After putting all this together, we find
\begin{equation}
\biggl< \biggl(\vv (\vx_1) \cdot \vn_1 \biggr) 
            \biggl( \vv (\vx_2) \cdot \vn_2\biggr) \biggr> = 
 \sum_l \frac{2l+1}{4\pi} \;C^{vv}_l\; P_l(\ctonetwo )
\label{eq:corv5}
\end{equation}

where the power spectrum multipoles $C^{vv}_l$ are given by
\begin{equation}
C^{vv}_l =  \frac{4\pi}{2l+1} \bigl(l{\cal B}_{l-1} + (l+1){\cal B}_{l+1} 
            \bigr)
\label{eq:cls}
\end{equation}

and the ${\cal B}_l$'s are defined as

\begin{equation}
{\cal B}_l \equiv  4\pi \int \frac{k^2dk}{(2\pi)^3} P_{vv}(k)\; j_l(k x_1)
        j_l(k x_2)
\label{eq:als}
\end{equation}

\vspace{1.cm}

\section{Appendix C}
\vspace{.4cm}

\noindent {\bf The {\em all sky} kSZ correlation function}
\vspace{.4cm}

Using eq.(\ref{eq:dtksz2}), we can write the average product of kSZ temperature
anisotropies along two directions $\vn_1$, $\vn_2$ as
\[
\langle \frac{\delta T_{kSZ}}{T_0}(\vn_1) \frac{\delta T_{kSZ}}{T_0}(\vn_2)
\rangle = \int \;dr_1dr_2dM_1dM_2\;d\vy_1d\vy_2
 \dot{\tau}(M_1) \dot{\tau}(M_2)\;W^{gas}(\vy_1-\vrv_1) 
W^{gas}(\vy_2-\vrv_2) \; \times
\]
\begin{equation}
\phantom{xxxxxxxxxxxx}\langle n(M_1,\vy_1) 
 \biggl( \frac{\vv (\vy_1)}{c}\cdot \vn_1\biggr) n(M_1,\vy_2) 
 \biggl( \frac{\vv (\vy_2)}{c}\cdot \vn_2\biggr) \rangle
\label{eq:proddtksz}
\end{equation}
Now we recall our model for the number of haloes of eq.(\ref{eq:nclcond})
to rewrite the ensemble average in eq.(\ref{eq:proddtksz}) as
\[
\langle n(M_1,\vy_1) 
 \biggl( \frac{\vv (\vy_1)}{c}\cdot \vn_1\biggr) n(M_1,\vy_2) 
 \biggl( \frac{\vv (\vy_2)}{c}\cdot \vn_2\biggr) \rangle \;\;= \;\;
 {\bar n}(M_1,z_1) \; \delta^D(M_1-M_2)\; \delta^D(\vy_1-\vy_2) 
                                      \;\; \sigma_{vv}^2(M_1)\; +
\]
\begin{equation}
 {\bar n}(M_1,z_1){\bar n}(M_2,z_2) \biggl< \biggl( \frac{\vv (\vy_1)}{c}\cdot \vn_1\biggr)
\biggl( \frac{\vv (\vy_2)}{c}\cdot \vn_2\biggr) \biggr> \; +
\parone\partwo \frac{\langle \phi(\vy_1,\vn_1) \phi(\vy_2,\vn_2) \rangle}{c^2}
\label{eq:mainproduct}
\end{equation}
The first term is the Poisson term, and is zero unless both $\vn_1$ and 
$\vn_2$ are looking at the same cluster. The second term is the 
velocity-velocity ({\it vv}) term, and the third contains the coupling
of density and velocity studied in Appendix A. Plugging 
eq.(\ref{eq:mainproduct}) into eq.(\ref{eq:proddtksz}) and writing the
integrands in terms of integrals in Fourier domain, one finds
\[
\langle \frac{\delta T_{kSZ}}{T_0}(\vn_1) \frac{\delta T_{kSZ}}{T_0}(\vn_2)
\rangle =
\int \;dr_1dr_2dM_1\;d\vy_1\; 
\dthreek \dthreeq \;\;e^{-i\;\vk(\vrv_1-\vy_1)+i\;\vq(\vrv_2-\vy_1)}
\; W^{gas}_k (W^{gas}_q)^*\dot{\tau}^2(M_1) {\bar n} (M_1,z_1)\;
                 \frac{\sigma_{vv}^2(M_1)}{c^2} \;\;\;+ 
\]
\[
\phantom{x} 
\int \;dr_1dr_2dM_1dM_2\; 
\dthreek \dthreeq \;\;e^{-i\;\vk\cdot\vrv_1+i\;\vq\cdot\vrv_2}
{\bar n}(M_1,z_1){\bar n}(M_2,z_2)\;W^{gas}_k(W^{gas}_q)^*\;\dot{\tau}(M_1)\dot{\tau}(M_2)
\biggl< \frac{\vv_{\vk}\cdot \vn_1}{c}\frac{\vv_{\vq}^*\cdot \vn_2}{c} 
   \biggr>  \;\;\;+ 
\]
\begin{equation}
\int \;dr_1dr_2dM_1dM_2\; 
\dthreek \dthreeq \;\;e^{-i\;\vk\cdot\vrv_1+i\;\vq\cdot\vrv_2}
\parone\partwo\;W^{gas}_k(W^{gas}_q)^*
 \dot{\tau}(M_1)\dot{\tau}(M_2) \;
\biggl< \frac{\phi_{\vk,\vn_1} \phi_{\vq,\vn_2}^*}{c^2} \biggr>. 
\label{eq:dtksz_f1}
\end{equation}
The integration on $\vy_1$ in the Poisson term generates a Dirac delta 
on $\vk-\vq$, whereas in the other two terms this Dirac delta arises
naturally when one computes the ensemble average product of the Fourier
modes of $v$ and $\phi$. Hence, the integral on $\vq$ dissappears 
and every integral ends up with a term of the form
$\exp{i\vk\cdot(\vrv_1-\vrv_2)}$. We introduce now the Rayleigh 
expansion, yielding
\begin{equation}
\exp{i\vk\cdot(\vrv_1-\vrv_2)} = \sum_{l,l'} (2l+1)(2l'+1) (-i)^{l-l'} \;
\; j_l(kr_1) j_{l'}(kr_2)\; 
P_l(\mu_{\vk,\vrv_1})P_{l'}(\mu_{\vk,\vrv_2}),
\label{eq:raydiff}
\end{equation}
 and we make use (as in  eq.(\ref{eq:corv4}) in Appendix B) 
of the addition theorem of Legendre
polynomia to express those Legendre polynomia having as 
argument $\mu_{\vk,\vrv_2}$ as a sum of Legendre functions of 
$\mu_{\vk,\vrv_1}$ and $\mu_{\vrv_1,\vrv_2}$. Here, as before, we have 
aligned the polar axis of $\hat{\vk}$ along $\vrv_1$ or $\vn_1$. 
Also in this case, the integral of the azymuthal angle of $\hat{\vk}$
cancels the contribution of all Legendre functions having $m\neq 0$,
so at the end one is left with
\[
\langle \frac{\delta T_{kSZ}}{T_0}(\vn_1) \frac{\delta T_{kSZ}}{T_0}(\vn_2)
\rangle = \sum_l \frac{2l+1}{4\pi}\; P_l( \cos \theta_{12})\;\; 
  \biggr[\;
\int \;\dthreekrad\;dr_1dr_2\;dM_1 \;\;
W^{gas}_k (W^{gas}_q)^*
 \; \dot{\tau}^2(M_1)\;
    \frac{\sigma_{vv}^2(M_1)}{c^2} \;j_l(kr_1)j_l(kr_2)\;\;+ 
\]
\[
\phantom{x} 
\int \;\dthreekrad\;dr_1dr_2dM_1dM_2\; {\bar n}(M_1,z_1){\bar n}(M_2,z_2)
\;W^{gas}_k(W^{gas}_q)^*\;\dot{\tau}(M_1)\dot{\tau}(M_2)\;\frac{P_{vv}(k)}{c^2}
\; 
j_l(kr_1)j_l(kr_2)\; \cos \theta_{12}\;\;\;+ 
\]
\[
\phantom{x} 
\int \;\dthreekrad\;dr_1dr_2dM_1dM_2\; 
\parone\partwo\;W^{gas}_k(W^{gas}_q)^* \;
\;  \dot{\tau}(M_1)\dot{\tau}(M_2) \;\bigl[P_{dd}\star P_{vv}\bigr](k)\;
j_l(kr_1)j_l(kr_2)\; \cos \theta_{12}\;\;\;+ 
\;\;
\]
\begin{equation}
\int \;\dthreekrad\;dr_1dr_2dM_1dM_2\; 
\; 
\parone\partwo\;W^{gas}_k(W^{gas}_q)^* \;
\;  \dot{\tau}(M_1)\dot{\tau}(M_2) \;\bigl[P_{dv}\star P_{vd}\bigr](k)\;
\;
j_l(kr_1)j_l(kr_2)\; \biggl(\sin \theta_{12} + \cos \theta_{12} \biggr)
     \biggr]
\label{eq:cfas_pf}
\end{equation}

As in Appendix A, we have taken $\vn_1 = (0,0,1)$ and 
$\vn_2=(0,\sin \theta_{12}, \cos \theta_{12})$. Since different velocity
components are not correlated, and we are only sensitive to the radial
projection of the velocity, a  $\cos{\theta_{12}}$ dependence appears 
in the {\it vv} term. For exactly the same reasons
we obtained a $\cos{\theta_{12}}$
and a $\cos{\theta_{12}} + \sin{\theta_{12}}$ dependence 
when computing the power spectrum of $\phi$ in Appendix A. Note that, out
of the three terms we found in that computation, we have 
dropped the last one because it is constant and introduces no anisotropy.
 Since the
power spectra multipoles ($C_l$'s) are projections on Legendre polynomia, such
projection must be applied on $\cos{\theta_{12}}$
and $\cos{\theta_{12}} + \sin{\theta_{12}}$. The projection matrix for
$\cos{\theta_{12}}$ will be identical to that given by eq.(\ref{eq:cls})
in Appendix B. For $\cos{\theta_{12}} + \sin{\theta_{12}}$, it must be computed
numerically. \\

Summarising, the power spectra corresponding to each of the
terms considered here can be understood as a transformation of some vectors
$c^X_{l'}$ (where $X$ runs for {\it Poisson, vv, dd-vv, dv-vd}) by some
linear applications ${\cal A}^X_{l,l'}$:
\begin{equation}
C_l^X = \sum_{l'} {\cal A}^X_{l,l'} c^X_{l'}.
\label{eq:transfcl}
\end{equation}
 For the Poisson term,
${\cal A}^{Poisson}_{l,l'}$ is the identity, whereas for the {\it vv} 
 and the {\it dd-vv} terms we find
\begin{equation}
{\cal A}_{l,l'}^{vv,\;dd-vv} = 4\pi\;\biggl[ 
\frac{l\;\delta_{l-1,l'}^K}{(2l-1)^2} + 
            \frac{(l+1)\;\delta_{l+1,l'}^K}{(2l+3)^2} 
\biggr],
\label{eq:A_cos_bis}
\end{equation}
with $\delta^K_{i,j}$ is the Kronecker delta for $i$ and $j$. The projection
matrix for the {\it dv-vd} term reads
\begin{equation}
{\cal A}_{l,l'}^{dv-vd} = 2\pi\;\; \int_{-1}^{+1}d\mu \;
          (\mu+\sqrt{1-\mu^2}) P_l (\mu) P_{l'} (\mu ).
\label{eq:A_cos_sin_bis}
\end{equation}

The $c_l^X$ 
vectors for the Poisson, {\it vv}, {\it dd-vv}, and {\it dv-vd} terms
are as follows:
\begin{itemize}
\item {\it Poisson}:

\begin{equation}
c^{Poisson}_l = \frac{2}{\pi} \int \dthreekr\;dM\; \bigr(
                 \Delta^P_l (k,M)\bigl)^2,
\label{eq:cl_poisson_bis}
\end{equation}
with $\Delta^P_l(k,M)$ being
\begin{equation}
\Delta^P_l(k,M) = \int dr \;\dot{\tau} \;\sigma_{vv} \bigl[ n(M,z)\bigr]^{1/2}
  j_l(kr) W_k^{gas}.
\label{eq:dl_poisson_bis}
\end{equation}

\item {\it vv} term:
\begin{equation}
c^{vv}_l = \frac{2}{\pi} \int \dthreekr\; 
                \biggl( \Delta^{vv}_l (k) \biggr)^2,
\label{eq:cl_vv_bis}
\end{equation}
with $(\Delta^{vv}_l(k) )^2$ given by
\begin{equation}
\biggr( \Delta^{vv}_l(k)\biggr)^2 = 
\int dr_1\;dr_2\;dM_1\;dM_2\;\taud(M_1,z_1)\taud(M_2,z_2)\;
\;\;n(M_1,z_1)n(M_2,z_2)\;W^{gas}_k(M_1,z_1)W^{gas}_k(M_2,z_2)
\; P_{vv}(M_1,M_2,z_1,z_2,k)\; j_l(kr_1)j_l(kr_2).
\label{eq:dl_vv_bis}
\end{equation}

\item {\it dd-vv} term:
\begin{equation}
c^{dd-vv}_l = \frac{2}{\pi} \int \dthreekr \biggl( \Delta_l^{dd-vv}
 \biggr)^2
\label{eq:cl_dd_vv_bis}
\end{equation}
with
\[
\biggr( \Delta^{dd-vv}_l(k)\biggr)^2 = 
\int dr_1\;dr_2\;dM_1\;dM_2\;\taud(M_1,z_1)\taud(M_2,z_2)\;
\;\;\parone\partwo\;W^{gas}_k(M_1,z_1)W^{gas}_k(M_2,z_2) \;\; 
\times
\]
\begin{equation}
\phantom{xxxxxxxxxxxxxxxxxxxxxxxx} [P_{dd}\star P_{vv}](M_1,M_2,z_1,z_2,k)\; 
                         j_l(kr_1)j_l(kr_2).
\label{eq:dl_dd_vv_bis}
\end{equation}

\item {\it dv-vd} term:

\begin{equation}
c^{dv-vd}_l = \frac{2}{\pi} \int \dthreekr \biggl( \Delta_l^{dv-vd}
 \biggr)^2;
\label{eq:cl_dv_vd_bis}
\end{equation}
with
\[
\biggr( \Delta^{dv-vd}_l(k)\biggr)^2 = 
\int dr_1\;dr_2\;dM_1\;dM_2\;\taud(M_1,z_1)\taud(M_2,z_2)\;
\;\parone\partwo\;W^{gas}_k(M_1,z_1)W^{gas}_k(M_2,z_2)
\;\;\times
\]
\begin{equation}
\phantom{xxxxxxxxxxxxxxxxxxxxxxxxxxxxxxxxxx} 
    [P_{dv}\star P_{vd}](M_1,M_2,z_1,z_2,k)\; 
                         j_l(kr_1)j_l(kr_2).
\label{eq:dl_dv_dv_bis}
\end{equation}

\end{itemize}

\label{lastpage}




\end{document}